\documentclass[a4paper,prd,twocolumn,showpacs,superscriptaddress,floatfix,nofootinbib]{revtex4-1}
\usepackage{graphicx}
\usepackage{amsmath}
\usepackage{amssymb}
\usepackage{pifont}
\usepackage{dcolumn}
\usepackage{bm}
\usepackage{enumitem}   
\usepackage{xcolor}
\usepackage{svg}
\usepackage{natbib}
\usepackage{comment}
\usepackage{hyperref}
\usepackage{xspace}
\hypersetup{
    colorlinks=true,
    linkcolor=blue,
    citecolor=blue,
    filecolor=magenta,      
    urlcolor=cyan
    }
\usepackage{times}

\newcommand{\eg}{e.g.,~}
\newcommand{\ie}{i.e.,~}

\newcommand{\orcid}[1]{\href{https://orcid.org/#1}{
\includegraphics[width=10pt]{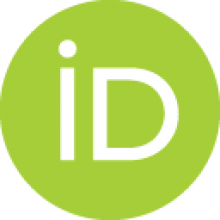}}}
\newcommand{\BHAC}{\texttt{BHAC+}\xspace}
\newcommand{\BHACOLD}{\texttt{BHAC}\xspace}
\newcommand{\FIL}{\texttt{FIL}\xspace}
\newcommand{\FUKA}{\texttt{FUKA}\xspace}
\newcommand{\HO}{hand-off\xspace}
\newcommand{\GWBR}{RR\xspace}

\begin{document}

\title{Long-term impact of the magnetic-field strength on the evolution
  and electromagnetic emission by neutron-star merger remnants}

\author{Jin-Liang Jiang\:\orcid{0000-0002-9078-7825}}
\affiliation{Institut f\"ur Theoretische Physik, Goethe Universit\"at,
  Max-von-Laue-Str. 1, 60438 Frankfurt am Main, Germany}

\author{Harry Ho-Yin Ng\:\orcid{0000-0003-3453-7394}}
\affiliation{Institut f\"ur Theoretische Physik, Goethe Universit\"at,
  Max-von-Laue-Str. 1, 60438 Frankfurt am Main, Germany}

\author{Michail Chabanov\:\orcid{0000-0001-9676-765X}}
\affiliation{Center for Computational Relativity and Gravitation \&
  School of Mathematical Sciences, Rochester Institute of Technology, 85
  Lomb Memorial Drive, Rochester, New York 14623, USA}
\affiliation{Institut f\"ur Theoretische Physik, Goethe Universit\"at,
  Max-von-Laue-Str. 1, 60438 Frankfurt am Main, Germany}

\author{Luciano Rezzolla\:\orcid{0000-0002-1330-7103}}
\affiliation{Institut f\"ur Theoretische Physik, Goethe Universit\"at,
  Max-von-Laue-Str. 1, 60438 Frankfurt am Main, Germany}
\affiliation{School of Mathematics, Trinity College, Dublin 2, Ireland}
\affiliation{Frankfurt Institute for Advanced Studies,
  Ruth-Moufang-Str. 1, 60438 Frankfurt am Main, Germany}

\date{\today}

\begin{abstract}
Numerical simulations are essential to understand the complex physics
accompanying the merger of binary systems of neutron stars. However,
these simulations become computationally challenging when they have to
model the merger remnants on timescales over which secular phenomena,
such as the launching of magnetically driven outflows, develop. To tackle
these challenges, we have recently developed a hybrid approach that
combines, via a hand-off transition, a fully general-relativistic code
(\FIL) with a more efficient code making use of the conformally flat
approximation (\BHAC). We here report important additional developments
of \BHAC consisting of the inclusion of gravitational-wave
radiation-reaction contributions and of higher-order formulations of the
equations of general-relativistic magnetohydrodynamics. Both improvements
have allowed us to explore BNS merger remnants with high accuracy and
over timescales that would have been computationally prohibitive
otherwise. More specifically, we have investigated the impact of the
magnetic-field strength on the long-term (\ie $\sim 200\,{\rm ms}$) and
high-resolution (\ie $150\,{\rm m}$) evolutions of the ``magnetar''
resulting from the merger of two neutron stars with a realistic equation
of state. In this way, and for sufficiently large magnetic fields, we
observe the weakening or suppression of differential rotation and the
generation of magnetic flares in the outer layers of the remnant. These
flares, driven mostly by the Parker instability, are responsible for
intense and collimated Poynting flux outbursts and mass ejections. This
novel phenomenology offers the possibility of seeking corresponding
signatures from the observations of short gamma-ray bursts and hence
revealing the existence of a long-lived strongly magnetized remnant.
\end{abstract}
\maketitle

\section{Introduction}
\label{sec:intro}

The observation of the gravitational-wave (GW) event GW170817 and the
corresponding electromagnetic (EM) emission~\cite{Abbott2017,
  Abbott2017_etal, Abbott2017b} marked the dawn of a new era in
multi-messenger astronomy, combining GW and EM observations. This
breakthrough in multi-messenger signals provides us with unprecedented
opportunities to address long-standing questions in physics, such as the
behavior of the equation of state (EOS) for extremely dense nuclear
matter, the mechanisms behind the launching of relativistic jets and
short gamma-ray bursts (GRBs)~\cite{Rezzolla:2011, Just2016, Ciolfi2020,
  Hayashi2021, Baiotti2016, Paschalidis2016, Murguia-Berthier2016,
  Lazzati2017c}, and the origin of heavy elements in the
universe~\cite{Metzger:2010, Bovard2017, Smartt2017, Papenfort2018,
  Combi2022, Fujibayashi2023, Kawaguchi2023}. To answer these questions
thoroughly, it is necessary to model the highly nonlinear processes
involved in these multi-messenger events with accuracy and realism. This
requires self-consistent numerical simulations that solve the Einstein
equations, general-relativistic magnetohydrodynamics (GRMHD), and
neutrino radiation transport, while incorporating realistic tabulated
EOSs. These advanced techniques are vital for capturing the intricate
details of binary neutron-star (BNS) merger dynamics and bridging
theoretical predictions with observational data.

Although significant progress has been achieved in numerically modelling
BNS mergers in recent years, two major aspects still require further
investigation. The first pertains to high-resolution simulations that
focus on MHD processes, such as the Kelvin-Helmholtz instability
(KHI)~\cite{Rasio99, Price06, Anderson2008, Baiotti08, Chabanov2022}, the
$\alpha$-$\Omega$ dynamo~\cite{Ruediger1993, Bonanno:2003uw, Most2023b,
  Musolino2024b}, the magnetorotational instability
(MRI)~\cite{Velikhov1959, Chandrasekhar1960, Chandrasekhar1960,
  Siegel2013}, and magnetic winding/braking~\cite{Siegel2014, Kiuchi2017,
  Kiuchi2022}, which may all occur during BNS mergers. The impact of
these MHD processes on the evolution of BNS mergers is enormous, but hard
to quantify as several of them are highly sensitive to grid
resolution. For instance, the MRI contributes to the exponential growth
of large-scale poloidal magnetic fields, facilitating angular momentum
transfer, influencing black-hole collapse time, driving strong postmerger
winds, and potentially affecting relativistic jet formation. However, the
MRI's effectiveness depends sensitively on simulation resolution and on
the magnetic-field topology generated by the KHI process, which is also
highly resolution dependent. Given the critical dependence on resolution
for both instability processes, and the lack of convergence in current
simulations, there is an urgent need for high-resolution
simulations. Yet, these simulations are computationally expensive and
time-intensive, necessitating efficient approaches to reach feasible
solutions.

The second aspect involves an urgent need for long-term, large-scale
simulations to consistently capture all multi-messenger signals produced
in compact binary merger events, \eg GWs~\cite{LIGOScientific:2017vwq,
  LIGOScientific:2018hze}, GRB multi-band EM emission~\cite{FermiLat2017,
  Goldstein2017}, multi-band afterglow emission~\cite{Hallinan2017,
  Troja2017, Margutti2018, Mooley2018, Ghirlanda2019, Musolino2024}, and
kilonova observations~\cite{Chornock_etal2017, Li1998, Drout2017,
  Nicholl2017, Pian2017, Soares-Santos2017}. These general-relativistic
simulations must cover timescales of seconds to account for the delay
from the GW to the associated GRB signal, which is $1.7\,\mathrm{s}$ in
the case of GW170817~\cite{Goldstein2017}, and hints at a collapse time
of about $1.0\,\mathrm {s}$~\cite{Gill2019, Murguia-Berthier2020,
  Nathanail2021}. Even longer timescales, although not necessarily in
full general relativity (GR), must be span to describe the afterglow
signal appearing days later~\cite{Margutti2017} and lasting
months~\cite{Troja2019}. This presents challenges in maintaining the same
simulation method over extended periods. Hence, transitions between codes
(\HO) at different stages offer a versatile approach, enabling various
physical processes to be modelled while leveraging the unique strengths
of different codes (see, \eg~\cite{Armengol2022, Gottlieb:2023a,
  Gottlieb:2023b, Gottlieb:2023c, Ng2024b, Ennoggi2025}).

Using the Conformal Flatness Condition (CFC) to solve the elliptic sector
of the Einstein equations offers an efficient approach to these
challenges, as it reduces the need for frequent spacetime computations
when the spacetime is not highly dynamic~\cite{Ng2024b}. The CFC
approximation has been successfully applied in simulations of
core-collapse supernovae~\cite{Dimmelmeier02b, Ott07b, Muller2015,
  Cheong2023, Ng2024a}, rapidly rotating neutron
stars~\cite{Dimmelmeier02a, Cordero2009, Ng2021, Cheong2024}, and BNS
mergers~\cite{Bauswein2012, Bauswein2020c, Lioutas2024}. Our previous
work~\cite{Ng2024b}, further demonstrated that the extended-CFC (xCFC)
scheme~\cite{Cordero2009} implemented in our new , multi-coordinate and
multi-dimensional GRMHD code \BHAC can effectively handle the output of
another code, \FIL~\cite{Most2019b, Most2020e, Chabanov2022,
  Musolino2023, Chabanov2023, Ng2024c}, which solves the Einstein
equations using full general relativity. Specifically, we have shown that
the CFC approximation provides nearly identical solutions in the head-on
collision of two neutron stars, not only in the evolution of maximum
rest-mass density, but also in the two-dimensional (2D) rest-mass density
and temperature distributions, when compared to the full-GR code \FIL.

In this work, we present the implementation, testing and use of three
major extensions to \BHAC. First, we go beyond the limitations of CFC by
introducing a phenomenological GW radiation-reaction (RR) scheme into
\BHAC~\cite{Ng2024b}. This addition compensates for the effects of
residual GW emission and enables more accurate modelling of the fluid and
magnetic-field dynamics during the postmerger phase. Our calculations of
the RR corrections are based on Refs.~\cite{Faye2003, Oechslin07a}, with
significant improvements to both the accuracy of the scheme and its
coupling to the xCFC framework. Second, we have improved several
numerical algorithms to higher-order ones, such as replacing the
second-order finite-volume method with a fourth-order finite-difference
approach, and switching from the Piecewise Parabolic Method (PPM)
reconstruction~\cite{Colella84} to a WENO-Z+
reconstruction~\cite{Acker2016}. Third, the \HO method has been extended
for parallel execution and optimised significantly by reducing the memory
requirements.

Leveraging on these improvements, and exploiting the results of
simulations of the inspiral and merger of BNSs computed by the full-GR
code \FIL, we have conducted with \BHAC long-term and three-dimensional
simulations of the merger remnant, starting from different postmerger
stages to verify the consistency between the results of \FIL and
\BHAC. We found that the xCFC scheme remains highly accurate, robust, and
efficient in capturing the long-term postmerger evolution of the fluid
and magnetic fields. This holds true despite the differences in numerical
methods for evolving spacetime and handling the divergence-free condition
between \BHAC~and \FIL.

Using this approach, we were able to perform four long-term simulations of
neutron-star merger remnants differentiating either in the strength of
the magnetic field (high/low) and in the resolution ($300/150\,{\rm
  m}$). These simulations have allowed us to assess the impact of the
magnetic-field strength on the evolution and EM emission by neutron-star
merger remnants, revealing, for instance, that very strong magnetic
fields can moderate or suppress the differential rotation in the
remnant. Additionally, they can lead to the generation of a strong and
collimated Poynting flux after the Parker instability criterion is met
and to the generation of magnetic flares in the outer layers of the
remnant that could accompany the phenomenology of short GRBs.

The paper is organised as follows. In Sec.~\ref{sec:math_setup}, we
describe the mathematical formulation of the CFC approximation and
outline the coupling of a novel RR module to the xCFC scheme. The basic
numerical setup and the details of the \HO are presented in
Sec.~\ref{sec:num_setup}, while in Sec.~\ref{sec:results}, we validate
the \HO and RR through a series of simulations. We focus on the long-term
impact of strong magnetic fields on the merger remnant in
Sec.~\ref{sec:long_term}. Finally, we discuss our results in
Sec.~\ref{sec:summary} and suggest possible directions for future
research.

\noindent Throughout this paper, unless stated otherwise, we adopt units
in which $c = G = M_{\odot} = k_{\rm B} = \epsilon_0 = \mu_0 = 1$, except
for the coordinates. Latin indices denote spatial components (from $1$ to
$3$), while Greek indices indicate spacetime components (from $0$ to
$3$).

\section{Mathematical Setup}
\label{sec:math_setup}

\subsection{The CFC approximation and extended CFC scheme}
\label{sec:grmhd}

We recall that the CFC reduces the conformally-related spatial metric
$\tilde{\gamma}_{ij}$ to the flat metric $f_{ij}$. Thus, the spatial
metric can be expressed in terms of the conformal factor $\psi$ and the
flat metric, \ie
\begin{equation}
  \label{eq:cfc}
  \gamma_{i j}=\psi^4 \tilde{\gamma}_{i j}\,.
\end{equation}
This approximation suppresses the radiative degrees of freedom in the
Einstein equations and is thus known as the ``waveless
approximation''. By employing the maximal slicing condition $K = \gamma^
{ij} K_{ij} = 0$, where $K_{ij}$ is the extrinsic curvature, the
Hamiltonian and momentum-constraint equations reduce to a set of coupled
nonlinear elliptic differential equations. This set of elliptic equations
can be further refined into a well-established scheme known as the xCFC
scheme~\cite{Cordero2009}.

In the context of the xCFC scheme implemented in \BHAC~\cite{Ng2024b},
the resulting set of elliptic equations is given by
\begin{align}
  &\hat{\Delta} X^i+\frac{1}{3}
  \hat{\nabla}^i\left(\hat{\nabla}_j X^j\right)=8 \pi f^{i j}
  \tilde{S}_j\,,
  \label{eq:xcfc_eq_x} \\
  &\hat{\Delta} \psi=-2 \pi \psi^{-1} \tilde{E} -\frac{1}{8} \psi^{-7}f_{i k} f_{j l}
  \hat{A}^{k l} \hat{A}^{i j} \,,
  \label{eq:xcfc_eq_psi} \\
  &\hat{\Delta}(\alpha \psi)=(\alpha \psi)\left[2 \pi \psi^{-2}(\tilde{E}+2
    \tilde{S}) + \frac{7}{8} \psi^{-8} f_{i k} f_{j l} \hat{A}^{k l} \hat{A}^{i j}
    \right]\,, \label{eq:xcfc_eq_alpha} \\
  &\hat{\Delta} \beta^i+\frac{1}{3} \hat{\nabla}^i\left(\hat{\nabla}_j
  \beta^j\right)=16 \pi \alpha
  \psi^{-6} f^{i j} \tilde{S}_j+2 \hat{A}^{i j}
  \hat{\nabla}_j\left(\psi^{-6} \alpha\right)\,, \label{eq:xcfc_eq_beta}
\end{align}
where $\tilde{S}_j := \psi^6 S_j$, $\tilde{E} := \psi^6 E$, and $S$ and
$E$ are the conserved momentum flux and the conserved energy density,
respectively. The tensor $\hat{A}^{ij}$ is the conformal extrinsic curvature, given by
\begin{equation}
\hat{A}^{i j} \approx \hat{\nabla}^i X^j+\hat{\nabla}^j X^i-\frac{2}{3}
\hat{\nabla}_k X^k f^{i j}\,.
\label{eq:xcfc_eq_aij}
\end{equation}
(for more details on the mathematical formulation and technical aspects,
we refer the reader to Refs.~\cite{Cordero2009, Cheong2020, Cheong2021,
  Ng2024b}).

A crucial aspect of the CFC approach that makes it computationally
attractive is that the xCFC equations are not necessarily updated at
every time-step or sub-step of the Runge-Kutta method. Instead, we can
solve this set of elliptic equations at much lower frequency, which we
quantify with the ratio $\chi := \Delta t_{\rm met}/\Delta t_{_{\rm
    MHD}}$, where $\Delta t_{\rm met}$ is the time interval between two
metric updates, and $\Delta t_{_{\rm MHD}}$ is the time-step of the GRMHD
simulation. The value of $\chi$ depends on the resolution of the
simulation, the timescale of spacetime changes, and the degree of
conformal flatness, but can easily reach values $\chi \simeq 10$ and even
be as large as $\chi\simeq 40$ at later stages in the postmerger
evolution (see Appendix~\ref{sec:cfc_PM_update_freq} for a discussion on
the impact of different values of $\chi$ and the postmerger evolution).
The actual solution of the xCFC equations is made employing an efficient
and low-memory cell-centered multigrid solver~\cite{Cheong2020,
  Cheong2021}, as a result, the impact of the solution of the elliptic
sector is only $\sim 40\%$ of the total cost of a reference time-step
(similarly, the solution of the full Einstein equations represents $\sim
40\%$ of the time-step from \FIL). In addition, when solving the set of
GRMHD equations, the time-step size needs not be CFL-limited by the speed
of light, but by the relevant MHD speeds, which are $\sim 30\%$
smaller. As a result, speed-ups of the order of $3-4$ are seen in our
hybrid approach (see also Appendix~\ref{sec:cfc_PM_update_freq} and
Tab.~\ref{tab:cost} for a detailed discussion).

\subsection{GW radiation-reaction in the xCFC scheme}

Some of the initial work on the inclusion of RR on the energy and
momentum budget in numerical codes solving the post-Newtonian (PN)
formulation of the hydrodynamic equations was reported in
Refs.~\cite{Blanchet90, Rezzolla:1999sz, Faye2003}. These approaches
were then coupled phenomenologically with the CFC scheme thereby
compensating for the waveless property of the CFC
approximation~\cite{Oechslin07a, Lioutas2024}. By implementing the
3.5~PN order formalism, but including only all 2.5PN terms and 3.5~PN
corrections related to the gravitational potential and its derivatives in
the metric components, the RR scheme provides a gravitational description
with modified metric components coupled to hydrodynamics to approximate
the system's energy-momentum loss via GWs.

However, the previous approaches have several disadvantages and require
various approximations (see, \eg Ref~\cite{Oechslin07a}). First of all,
loss of accuracy can result if the time-step is not sufficiently
small. This is the result of the need to perform two additional
time-derivatives of the first time-derivative of the quadrupole moment
$I^{[1]}_{ij}$ which are computed using finite-differencing as outlined
in Ref.~\cite{Oechslin07a} (hereafter, $X^{[k]}$ denotes the $k$-th order
time-derivative of $X$ with respect to the coordinate time). To maintain
accuracy and stability, particularly when there are rapid changes in the
quadrupole moment $I_{ij}$, time-derivatives must be updated at every
sub-step of the time integration and with a very small
time-step. Secondly, because GW emission remains significant at $\sim
10-20\, \rm{ms}$ after merger (see also Fig.~\ref{fig:gw_ho}), high-order
time derivatives of the Newtonian mass quadrupole are not small and need
to be properly accounted for. Finally, previous work (\eg
Refs.~\cite{Oechslin07a, Lioutas2024}) has not provided a close
comparison between a CFC scheme including RR and a full-GR scheme, making
it hard to judge the validity of some of the assumptions made in
implementing the RR terms. To the best of our knowledge, this is the
first time that a detailed comparison is presented on the role of the RR
terms in the CFC approximation. Given the intricacies and subtleties in
computing these terms and those discussed above, we hope that this
information will be of use for future studies employing the CFC
approximation (and the RR corrections) in BNS postmerger simulations.

In view of these considerations, we have taken a fresh new look at the
whole approach of including RR terms within the 3.5~PN order formalism
introduced in Ref.~\cite{Oechslin07a}, and hence derived a modified RR
scheme. In this approach, we employ terms based on Ref.~\cite{Faye2003},
but include the magnetic energy in the estimation of the matter sources,
and couple the scheme to the metric components of the xCFC equations. We
therefore solve the following set of nonlinear elliptic equations in
terms of the six scalar potentials $\mathcal{U}_*, \mathcal{U}_{* i},
\mathcal{R}$, and $\mathcal{R}_2$ (we here consider $c=1$ but we report
it explicitly in Appendix~\ref{sec:GWBRderivation} to keep track of the
various PN orders)
\begin{subequations}
\begin{align}
& \Delta \mathcal{U}_*=-4 \pi \sigma\,, \label{eq:PNcorr1} \\
& \Delta \mathcal{U}_{* i}=-4 \pi \sigma w_i\,, \label{eq:PNcorr2}\\
& \Delta \mathcal{R}=-4 \pi Q_{i j}^{[3]} x^i \partial_j \sigma\,, \label{eq:PNcorr3}\\
& \Delta \mathcal{R}_2=-4 \pi \sigma\left(Q_{i j}^{[3]} x^i 
  \partial_j \mathcal{U}_* - 3 Q_{i j}^{[3]} w_i w_j - \mathcal{R}\right)\,, \label{eq:PNcorr4}
\end{align}
\end{subequations}
where $\sigma := T^{ii} + T^{00}$ is the 1~PN + 3.5~PN mass density to
replace the conserved rest-mass density $D^* := D \psi^6 = \rho W
\psi^6$~\cite{Oechslin07a} (see Eq.~\eqref{eq:sigma} in
Appendix~\ref{sec:GWBRderivation}), given that $T^{\mu \nu}$ is the
energy-momentum tensor, and $w_i := h u_i$ is the momentum per unit
rest-mass in the fluid frame as defined in Eq.~(4.2)
of~\cite{Blanchet90}. Furthermore, $u_\mu$ is the covariant fluid
four-velocity, and $h(\rho, T, Y_p) = 1 + \epsilon + (p+b^2)/\rho$ is the
total relativistic specific enthalpy, thus including the contribution
from EM fields ($b^2 = b^\mu b_\mu$ is the square of the magnetic field
strength in the fluid frame) in addition to that from the specific
internal energy $\epsilon$, the fluid pressure $p$, rest-mass density
$\rho$ and proton fraction $Y_p$.

Equations~(\ref{eq:PNcorr1})-(\ref{eq:PNcorr4}) correspond to
Eqs.~(4.31j), (4.31k), (4.31n), and (4.31cc) in~\cite{Faye2003}, where we
keep all 2.5~PN quantities and 3.5~PN corrections related to the PN
potential $\mathcal{R}_2$ and its derivatives, which depends on the
quantity $Q_{ij}^{[3]}$ [see Eq.~\eqref{eq:Q3}]. Note also that
Eq.~\eqref{eq:PNcorr4} includes the term $\sim 3 Q_{ij}^{[3]} w_i w_j$
that was omitted in Ref.~\cite{Oechslin07a} but that actually leads to
significant improvements in the accuracy of the RR scheme [see
  Appendix~\ref{sec:GWBRderivation} and the derivation of
  Eq.~(\ref{eq:R2})]. The quantity $Q_{i j}^{[3]}$ can be shown to be
related to the third time-derivative of the Newtonian mass quadrupole (we
will ignore here mass-current quadrupole but see
Ref.~\cite{Rezzolla:1999sz} for how to include these corrections), where
the two are related as~\cite{Blanchet90}
\begin{equation}
\label{eq:Q3}
  Q_{i j}^{[3]} = I_{ij}^{[3]} + \mathcal{O}\left(\frac{1}{c^2}\right)\,.
\end{equation}
Note that while $I_{i j}^{[3]}$ is a purely Newtonian (\ie 0~PN)
quantity, $Q_{i j}^{[3]}$ is a genuine PN quantity entering at 2.5~PN in
our scheme and computed as [see Eq.~(5.13) in~\cite{Blanchet90} and
Eq.~(B4) in~\cite{Faye2003}]

\begin{equation}
  \label{eq:Q3ij}
\begin{aligned}
Q_{i j}^{[3]}=& \Biggl[ \int_{\mathcal{V}} \sqrt{\hat{\gamma}} 
  d^3 x \, D^* \\ & 2\left[ w_i \partial_j \mathcal{U}_*- 2 w_i
    \left[\partial_j (h-1)-T \partial_j s\right] \right. \\ &+
    \left. x^i w_k \partial_{k} \partial_{j} \mathcal{U}_*-x^i
    \partial_{k} \partial_{j} \mathcal{U}_{*k}\right]\Biggr]^\mathrm{STF}\,,
\end{aligned}
\end{equation}
where $\mathcal{V}$ is the numerical domain, $s$ is the specific entropy
(entropy per baryon), 
$W$ the Lorentz factor, $\sqrt{\hat{\gamma}}$ the
determinant of the spatial part of the flat metric, and the superscript
${\rm STF}$ refers to a symmetric trace-free tensor, \ie $\left[A^{i
    j}\right]^{\rm STF}:=\frac{1} {2} (A^{i j} + A^{j i}) - \frac{1}{3}
\delta^{i j} A^{kk}$ denotes the symmetric trace-free part of the tensor.

We note that the use of $Q_{ij}^{[3]}$ computed as in Eq.~\eqref{eq:Q3ij}
avoids numerical time derivatives altogether and should be preferred to
the calculation of $I_{ij}^{[3]}$ from the actual second-order
time-derivative of the analytically calculated expression for
$I^{[1]}_{ij}$ proposed in Refs.~\cite{Oechslin07a, Lioutas2024} and
computed as in Ref.~\cite{Oechslin07a}
\begin{equation}
  \label{eq:I3ij}
\begin{aligned}
I^{[3]}_{ij}(t) =& 2 \left[ I^{[1]}_{ij}(t+dt_{1}) - [1+dt_{1}/dt_{2}]
  I^{[1]}_{ij}(t) +
  \right.\\ &\left. [dt_{1}/dt_{2}]I^{[1]}_{ij}(t-dt_{2}) \right]/ \left[
  dt_{1} (dt_{1}+dt_{2}) \right]\,,
\end{aligned}
\end{equation}
where $dt_{1}$ and $dt_{2}$ are the current and last time-steps,
respectively (we note that the calculation of $I^{[1]}_ {ij}$ itself
requires time-derivatives of several quantities, \eg $\rho$, $\psi$ and
$\epsilon$). By computing $Q_{i j}^{[3]}$ instead of $I_{i j}^{[3]}$, it
is possible to remove the source of potential numerical errors and the
numerous (and sometimes arbitrary) approximations made in their
computation, \ie the neglect of several terms in the total
time-derivatives of $\mathcal{U}$, $\partial_i \mathcal{U}$, $V^i$ and
$h$ in the expression of $I_{ij}$, where $\mathcal{U}$ and $V^i$ are the
Newtonian potential and the transport velocity,
respectively~\cite{Oechslin07a, Lioutas2024}\footnote{We note that some
of the approximations in Refs.~\cite{Oechslin07a, Lioutas2024} have been
made assuming that the neutron-star binaries are in the unrealistic
condition of corotation. However, these assumptions may be inaccurate in
the case of irrotational or spinning BNSs.}. In addition, our experience
has shown that the calculation of $I_{i j}^{[3]}$ is extremely inaccurate
near discontinuities or sharp gradients (\eg the stellar surface).
Reducing drastically the time-step can improve the calculation of $I_{i
  j}^{[3]}$ but obviously increases considerably the computational costs,
which we are instead interested in limiting. By contrast, computing $Q_{i
  j}^{[3]}$ is not only more accurate but also allows for a time-step
that is large and equal to that of the xCFC solver, \ie $\Delta t_{\rm
  met} = 0.5\, M_{\odot}$, chosen for postmerger phase.

By following Eqs.~(4.31a), (4.31b) in Ref.~\cite{Faye2003} and keeping
only the terms involving $Q_{ij}^{[3]}$, the metric component $g_{00}$
for the CFC system gains a new RR term given by
\begin{equation}\label{eq:g00_short}
g_{00, {_{\rm RR}}}=-\frac{4}{5}\left(1-2 \mathcal{U}_*\right)\left(Q_{i j}^{[3]} 
  x^i \partial_j \mathcal{U}_*-\mathcal{R}\right)+\frac{4}{5} \mathcal{R}_2\,.
\end{equation}
Note that the form of $g_{00, {_{\rm RR}}}$ is different from the
corresponding Eq.~(A.37) in~\cite{Oechslin07a} [or Eq.~(A.5) in
  Ref.~\cite{Lioutas2024}], which contains a typo~\cite{Oechslin2024} (we
refer the reader to the derivation presented in
Eqs.~(\ref{eq:g7})--(\ref{eq:g00}) in Appendix~\ref{sec:GWBRderivation}
for further details).

As a result, the new (primed) metric components modified by the inclusion
of RR take the form
\begin{align}
  \label{eq:g00_prime}
   g_{00}^{\prime} &:=  g_{00} + g_{00, {_{\rm RR}}} \,,
\end{align}
so that the new lapse function can be expressed as
\begin{align}
  \label{eq:alp_prime}
   \alpha^{\prime} &= \sqrt{\alpha^{2} - g_{00, {_{\rm RR}}} +
     \beta^{\prime i} \beta^{\prime}_i - \beta^i \beta_i }\,, \\
   & \approx \sqrt{\alpha^{2} - g_{00, {_{\rm RR}}}}\,,
\end{align}
where the second equality is obtained when assuming that $\beta^{\prime
  i} \beta^ {\prime}_i \simeq \beta^i \beta_i$, which is reasonable since
the leading order of $g_{0i, {_{\rm RR}}}$ is a 3.5~PN quantity and
$|\beta^ {\prime}_i - \beta_i| \ll 1$ in the postmerger phase.

In addition, we should remark that while there are RR corrections also in
the spatial part of the metric, \ie $g_{ij, {_{\rm RR}}} = \gamma_{ij,
  {_{\rm RR}}} = -(4/5)Q_{i j}^{[3]}$ [see Eqs.~(4.31d) and (4.31m) in
  Ref.~\cite{Faye2003}] these terms are difficult to incorporate in our
CFC approximation scheme. This is because of the intrinsically diagonal
form of the CFC equations, which allows us to replace $\psi$ with the
RR-corrected $\psi^{\prime}$, but does not give us access to the full
three-metric $\gamma_{ij}^{\prime}$. Using only the diagonal terms in
$\gamma_{ij, {_{\rm RR}}}$ is obviously mathematically inconsistent and,
unsurprisingly, when we have tried to apply $\psi^{\prime} =:
(\det{\gamma_{ij}^{\prime}})^{1/12}$ we have encountered significant
numerical problems such as non-preserved divergence-free condition and
less accurate conservation of rest-mass and energy-momentum. Hence, in
our implementation of the RR terms, we simply consider the components of
$\gamma_{ij, {_{\rm RR}}}$ as high-order PN terms and set them to
zero. To the best of our knowledge~\cite{Oechslin2024}, these terms are
neglected also in other implementations of the CFC scheme
(e.g.,~\cite{Oechslin07a}), although they can and are employed to compute
the GW signal.

Following Refs.~\cite{Blanchet90, Faye2003}, our coupling of the \GWBR
terms to the xCFC scheme is made using a three-dimensional (3D) Cartesian
coordinate system in which we compute the RR equations after every update
of the xCFC scheme. In particular, using a multigrid
algorithm~\cite{Cheong2021, Ng2024b, Lioutas2024}, we first solve the two
elliptic equations~(\ref{eq:PNcorr1}) and (\ref{eq:PNcorr2}) from which
we compute $Q_{i j}^{[3]}$ as in Eq.~(\ref{eq:Q3ij}). Next, we solve
Eqs.~(\ref{eq:PNcorr3}), (\ref{eq:PNcorr4}), and (\ref{eq:g00_short})
using the newly computed value for $Q_{i j}^{[3]}$. The full set of
elliptic equations (including
Eqs.~(\ref{eq:xcfc_eq_x})-(\ref{eq:xcfc_eq_beta})) is then solved
iteratively until the maximum absolute value of the residual falls below
a tolerance of $10^{-6}$, which is sufficient for convergence. A Robin
boundary condition is chosen for this procedure.

Once the correction $g_{00, {_{\rm RR}}}$ is computed from
Eq.~\eqref{eq:g00_short}, we obtain a modified lapse function
$\alpha^{\prime}$, which replaces $\alpha$ in the matter solver of the
GRMHD equations. However, when solving the xCFC metric equations, we
continue to use the original lapse function $\alpha$ instead. This
approach ensures that the RR impacts only the matter evolution and not
the field variables. We have found that, at least for the tests presented
here, not updating the lapse function for the xCFC equations does lead to
a better match with the results of the full-GR code \FIL. Additionally,
we also keep the conformal factor $\psi$ unchanged, thus ensuring that
metric modifications do not alter the conserved quantities on a given
spatial hypersurface. Not doing so would introduce modifications of the
three-metric $\gamma_{ij}$, thus compromising rest-mass conservation but
also the divergence-free condition for the magnetic field. Finally,
because Eq.~\eqref{eq:g00_short} does not guarantee positivity, it is
possible, under extreme conditions, that negative values of $g_{00,{_{\rm
      RR}}}$ appear in a small number of grid-points; in these cases we
simply set $g_{00,{_{\rm RR}}} = 0$.

\section{Numerical Setup}
\label{sec:num_setup}

\subsection{\FIL, \FUKA and \BHAC}
\label{sec:code_detail}

We here briefly recall the basic elements of our computational
infrastructure and report the relevant references for additional details.

\FIL~\cite{Most2019b, Most2020e, Chabanov2022, Musolino2023,
  Chabanov2023, Ng2024c} solves the hyperbolic sector of the Einstein
equations~\cite{Alcubierre:2006, Rezzolla_book:2013} using full GR
schemes, including BSSNOK~\cite{Shibata95, Baumgarte99},
BSSNOK-Z4c~\cite{Bernuzzi:2009ex}, and CCZ4~\cite{Alic:2011a,
  Alic2013}. Coupled with the
\texttt{EinsteinToolkit}~\cite{loeffler_2011_et,
  EinsteinToolkit_etal:2022_05}, it utilises the \texttt{Carpet}
box-in-box Adaptive Mesh Refinement (AMR) driver in Cartesian
coordinates~\cite{Schnetter:2006pg}. Besides, it uses a fourth-order
accurate conservative finite-difference High-Resolution Shock-Capturing
(HRSC) scheme with WENO-Z reconstruction~\cite{DelZanna2007}, an HLL
Riemann solver, and a vector-potential-based magnetic evolving
method~\cite{Etienne2015}.

The initial data of \FIL is generated using the publicly available
\texttt{FUKA} code~\cite{Grandclement09, Papenfort2021b, Tootle2023a},
which computes the initial data timeslice by solving the eXtended
Conformal Thin Sandwich system of equations~\cite{Pfeiffer:2005,
  Papenfort2021b}.

\BHAC~\cite{Ng2024b} (which is an extension of the publicly available
\BHACOLD~\cite{Porth2017, Olivares2019, Ripperda2019, Mpisketzis2024,
  Mpisketzis2024b}) solves the elliptic sector of the Einstein equations
using the xCFC scheme~\cite{Cordero2009}, with a cell-centered multigrid
solver~\cite{Cheong2020, Cheong2021} and block-based quadtree-octree
AMR~\cite{Xia2018, Keppens2021}. For consistency with \FIL, \BHAC is
extended to incorporate fourth-order finite-difference HRSC, the same
reconstruction method, and the same Riemann solver. However, it employs a
second-order convergent scheme for the xCFC metric solver (with or
without RR) and a magnetic-field-based upwind constrained-transport
method~\cite{Olivares2019}.

In the default setup, \BHAC makes use of the WENO-Z reconstruction and
fourth-order finite differencing and this leads to the closest matches
with the results from \FIL in terms of the evolution of maximum density,
rotational profiles, and total EM energy. It is important to mention that
both approaches to the violation of the divergence-free condition,
namely, the use of a vector potential or of the constrained-transport
method~\cite{Evans1988}, guarantee a magnetic field that is divergence
free at machine precision. It is also worth noting that we have resolved
an artefact reported in our previous work~\cite{Ng2024b}, where we
observed that when performing an azimuthally averaged \HO, the postmerger
remnant experienced fluctuating and significantly lower temperatures in
the core region. As discussed in~\cite{Ng2024b}, this behaviour was
partly due to slightly inaccurate values of the specific internal energy
$\epsilon$ generated by metric initialisation with the azimuthally
averaged \HO, but also to the poor resolution of the tabulated EOS, which
has a strong dependence of the temperature $T$ on the specific internal
energy $\epsilon$ in these high-density regimes. Although these artefacts
are reduced when employing a full 3D \HO, a series of experiments has
revealed that reconstructing the $T$ instead of the $\epsilon$ during
evolution leads to significant improvements of this issue and is
effective across various EOSs.

Finally, for the metric solver, the full-GR Z4c scheme in \FIL with the
\texttt{Antelope} code typically updates at every sub-step of the time
integrator, while in \BHAC, the metric solver is set to update every
$0.5~M_{\odot}$ for the postmerger simulations (see
Appendix~\ref{sec:cfc_PM_update_freq}). Both codes use a third-order
Runge-Kutta method for time integration and employ a
Courant-Friedrichs-Lewy (CFL) factor of $\mathcal{C}_{\rm CFL} =
0.2$. Unless otherwise specified, these parameters and numerical methods
are the default setups for both codes.

\subsection{Hand-off procedure of three-dimensional data}
\label{sec:handoff_detail}

As mentioned above, the \HO procedure is a key aspect of our approach and
allows us to leverage \FIL's accuracy during the inspiral phase and
\BHAC's efficiency, low-memory usage and AMR flexibility in the postmerger
phase. The \HO procedure was first presented in Ref.~\cite{Ng2024b}, but we
have here further refined it by employing a memory-efficient data-initialisation
method and by incorporating advanced higher-order interpolation schemes. Both
of these improvements are detailed below.

\subsubsection{Memory-efficient data initialisation}
\label{sec:initialization}

A non-trivial challenge to face when importing data from \FIL into \BHAC
in full-fledged and high-resolution AMR simulations is the different
strategies adopted by the two codes in handling the mesh refinements.
More specifically, while \FIL employs box-in-box refinement, \BHAC has
more flexibility of shaping the refinement boundaries offered by the
block-based quadtree-octree AMR. To cope with these differences, we adopt
a strategy to reshape the boundaries at each refinement level based on
values of the rest-mass density so that the boundary of a refinement
level is set by an iso-contour of $\rho$. This solution provides a natural
criterion reflecting the distribution of matter and leads to a
significant memory saving. Additionally, the finest refinement level is
shaped to encompass the dominant rest-mass contribution of the BNS
system. With this grid-structure adjustment, each data-point in \BHAC
must be interpolated from the original \FIL data.

Another aspect of the \HO we had to resolve was the reading procedure. In
particular, unlike other approaches where an intermediate restart file is
generated by a standalone script, \eg see~\cite{Armengol2022}, we
directly read and interpolate the data output from \FIL within \BHAC,
thereby reducing the amount of different software packages employed. This
represents a challenge especially in importing the high-resolution data,
as additional memory is required to store the new data from \FIL and
avoid frequent file-read operations. We recall that the data output from
\FIL consists of multiple \texttt{HDF5} files containing distinct patches
of the simulation domain and including the ghost zones. Hence, when
considering a given refinement block in \BHAC, we first determine the
number of \FIL \texttt{HDF5} files required to cover such a block in
\BHAC. Subsequently, only the relevant \FIL-output files are read and the
pertinent data are interpolated onto the \BHAC grid within each
refinement block. Besides, in order to save memory, which is essential in
high-resolution simulations, we de-allocate the data segments read from
\FIL after finishing initialisation of one refinement block and continue
with loading new pertinent data for the next refinement block. Finally, we
should also note that this process is performed in parallel on different
processors for different refinement blocks, which makes the \HO procedure
highly efficient. Overall, this method balances the file-reading
operations with the memory footprint of the \HO, and greatly benefits
from the parallel nature of this process.

\subsubsection{High-order interpolation for spacetime and matter quantities}

At \HO, the rest-mass density $\rho$, proton fraction $Y_p$, specific
internal energy $\epsilon$, Eulerian velocity $v^i$, and conformal factor
$\psi$ need to be interpolated from the gridpoint values in \FIL to those
in \BHAC that, as discussed above, do not coincide in general. The
conformal factor from \FIL is used to calculate the gauge-independent
conserved quantities, given by $\sqrt{\gamma / \hat{\gamma}}\left(D, S_j,
\tau, D Y_p, B^j\right)$. Once
Eqs.~(\ref{eq:xcfc_eq_x})--(\ref{eq:xcfc_eq_beta}) are solved, both the
metric and the primitive variables are updated~\cite{Ng2024b}. At the
same time, the gauge-independent conserved quantities are kept unchanged
during metric initialisation, and we refer readers to~\cite{Ng2024b} for
a detailed metric interpolation procedure. This approach has been shown
to reduce large initial perturbations in the \HO, which can otherwise
cause inconsistencies in the subsequent evolution~\cite{Ng2024b}.
Consequently, any differences in the primitive variables between \FIL and
\BHAC arise only from interpolation error and gauge
differences. Furthermore, after significant experimentation we have found
that the best choice to minimise differences in various global quantities
is to employ a third-order Lagrangian interpolation for the metric and
the fluid variables.

Special attention is required when importing the magnetic field, for
which two approaches are possible. We recall, in fact, that
while \FIL evolves the vector potential $A^{i}$, \BHAC solves the
induction equation directly in terms of the magnetic field components
$B^{i}$. Hence, it is in principle possible to either interpolate
directly the values of the magnetic field from \FIL as any other fluid
variable, or to actually interpolate the vector potential and then
compute the magnetic field in \BHAC by taking the curl operator. The
first approach preserves the magnetic-field structure but does not
guarantee it to be divergence-free. In contrast, the second method, while
introducing small structural differences in the magnetic field due to
metric disparities between \FIL and \BHAC, guarantees a divergence-free
field in \BHAC. We have obviously implemented both approaches and
concluded that importing the vector potential provides the most accurate
transfer of data. This is because the calculation of the magnetic field
from the vector potential yields a divergence-free constraint enforced at
machine precision and this is best suited for the constrained-transport
approach employed in \BHAC~\citep{Olivares2019} that benefits from data
that is as divergence-free as possible.

At the same time, since a Lagrangian interpolation does not guarantee
continuity in the first derivatives and since a curl operator needs to be
applied to the vector potential to derive the components of the magnetic
field, a different high-order interpolation approach needs to be employed
for the vector potential. Also in this case, after extensive
experimentation of different approaches and orders, we have concluded
that a third-order Hermitian interpolation for the vector potential
yields the needed accuracy and smoothness in the first derivatives at a
reasonable computational cost. 

\begin{table}
  \centering
  \footnotesize
  \begin{tabular}{lccccc}
    \hline 
    \hline 
    Tests &  Baryonic mass  & EM energy & Internal energy & \\
    \hline
    \texttt{PM-A}          &  $0.35\%$ & $1.97\%$    & $0.31\%$ \\
    \texttt{PM-B}          &  $0.29\%$ & $0.89\%$    & $0.28\%$ \\
    \texttt{PM-C}          &  $0.27\%$ & $0.85\%$    & $0.24\%$ \\
    \texttt{PM-D}          &  $0.21\%$ & $0.87\%$    & $0.20\%$ \\
    \hline 
    \hline 
  \end{tabular}
  \caption{Absolute values of the relative difference of various volume
    integrals in \BHAC and \FIL at different \HO stages. }
  \label{tab:global_diff}
\end{table}

The results of this admittedly elaborate interpolation procedure from
\FIL to \BHAC are summarised in Table \ref{tab:global_diff}, where we
compare the relative differences between \FIL and \BHAC in various
\textit{global} quantities computed via volume integrals, \ie the baryon
mass, the EM energy and the internal energy. Note also that the table
contains data relative to different \HO times as these also play a role
in determining the precision of the interpolation. This is because the
assumption of conformal flatness becomes increasingly more accurate as
the evolution proceeds (see also a discussion in Appendix~\ref{sec:cy})
so that the solutions in \FIL and \BHAC are very similar. Notwithstanding
the improvement that comes with hand-offs at later times, the largest
observed relative difference in baryon mass [see Eq.~\eqref{eq:bary_mass}
  for the definition] reported in Tab.~\ref{tab:global_diff} is
$0.35\%$. On the other hand, when considering a \textit{point-wise}
comparison between \FIL and \BHAC quantities we find that the large
majority of grid-points exhibit a relative difference of less than
$1\%$. Such relative differences can be larger in the presence of strong
shock waves, which are frequently formed in the less dense outer regions
of the merger remnant, but do not have an impact on the dynamics of the
high-density bulk flow. Indeed, the relative differences in the
high-density regions are $\lesssim 0.3\%$ for all \HO times and are
mostly due to differences in the field variables between \FIL and
\BHAC.

\begin{table}
  \centering
  \footnotesize
  \setlength{\tabcolsep}{0.2em} 
  \renewcommand{\arraystretch}{1.1}
    \begin{tabular}{lccccc}
    \hline 
    \hline 
    & HO time      & $|B^{\rm max}_0|$ (\FIL) & Res. (\FIL) & Res. (\BHAC) & $|h|^2$  \\
    & $[{\rm ms}]$ & $[{\rm G}]$ & $[{\rm m}]$ & $[{\rm m}]$ &  $[|h|^2_{\rm mer}]$ \\
    \hline
    short-term & & & & &\\
    \hline
    \texttt{PM-A}    & $~7.5$  & $3.7\times10^{15}$ & $300$ & $300$ & $1/4$  \\
    \texttt{PM-B}    & $13.5$  & $3.7\times10^{15}$ & $300$ & $300$ & $1/8$  \\
    \texttt{PM-C}    & $19.1$  & $3.7\times10^{15}$ & $300$ & $300$ & $1/16$ \\
    \texttt{PM-C-RR} & $19.1$  & $3.7\times10^{15}$ & $300$ & $300$ & $1/16$ \\
    \texttt{PM-D}    & $30.0$  & $3.7\times10^{15}$ & $300$ & $300$ & $1/32$ \\
    \texttt{PM-D-RR} & $30.0$  & $3.7\times10^{15}$ & $300$ & $300$ & $1/32$ \\
    \hline
    \hline
    long-term &  &  & &  \\
    \hline
    \texttt{LowB-LR}  & $20.0$   & $2.0\times10^{16}$ & $300$ & $300$ & $1/18$  \\
    \texttt{LowB-HR}  & $20.0$   & $2.0\times10^{16}$ & $300$ & $150$ & $1/18$  \\
    \texttt{HighB-LR} & $20.0$   & $1.0\times10^{17}$ & $300$ & $300$ & $1/18$  \\
    \texttt{HighB-HR} & $20.0$   & $1.0\times10^{17}$ & $300$ & $150$ & $1/18$  \\
    \hline
    \hline 
  \end{tabular}
  \caption{List of the simulations performed in this work. 
    In the short-term simulations, ``RR'' refers to evolutions including the RR
    corrections. In the long-term simulations, ``LR'' and ``HR'' denote
    the low and high-resolution cases, respectively. 
    HO time is the \HO time moment. Finally, the last
    column reports the fraction of the GW amplitude at which the \HO is
    performed.}
  \label{tab:ID}
\end{table}

\subsection{Initial data and simulation setups}
\label{sec:ID_num_setup}

The initial data is computed using the TNTYST~\cite{Togashi2017} EOS and
refers to equal-mass and irrotational binaries with a total ADM mass of
$2.55~M_ {\odot}$, initialised at a separation of $45\,\mathrm{km}$. When
the separation between the ``barycentres'' of the two stars is $\lesssim
13.3\,\mathrm{km}$, a purely poloidal magnetic field~\cite{Etienne2015,
  Chabanov2022} is seeded inside each neutron star. The seed magnetic
field is given by
\begin{equation}
  \label{eq:mag_seed}
  A_{i} =  A_{\rm b} \left[ -(x^j - x^j_{\rm NS}) \epsilon_{ij} \right] \,
  \max\{p - p_{\rm cut}, 0\}\,,
\end{equation}
for $i = x, y $ and $ A_z = 0 $, where $ A_{\rm b} $ is a constant chosen
to adjust the magnetic-field strength, $ \epsilon_{ij} $ is the
Levi-Civita symbol, $x_{\rm NS}$ denotes the coordinate center of the
neutron star, $p$ is the fluid pressure, and the cut-off pressure is set
to $p_{\rm cut} = 1.0 \times 10^{-8} $. Three different strengths for the
seeding magnetic field are used in this work: $3.7\times 10^{15}\, \rm
G$, $2.0\times 10^{16}\, \rm G$, and $1.0\times 10^{17}\, \rm G$. The
corresponding EM energies at the seeding time are $7.6\times 10^{47}$,
$2.0\times 10^{49}$, and $5.4\times 10^{50}\, \rm erg$, respectively. A
summary of the simulations performed in this work together with essential
information on the \HO is presented in Tab.~\ref{tab:ID}, where ``RR'' in
the simulation name denotes the simulation with RR scheme, while ``LR''
and ``HR'' refer to the low and high resolution cases, respectively.

Both codes adopt 3D Cartesian coordinates with $z$-symmetry and employ a
refinement ratio of $2$ with $6$ refinement levels, where the finest
level has a boundary at $24\,\mathrm{km}$ and a resolution of
$300\,\mathrm{m}$. For the simulations reported here, FIL has a
computational grid with outer boundaries at $1500\,\mathrm{km}$ in all
three spatial directions. In contrast, \BHAC adjusts the grid setup by
shrinking the outer boundary to $600\,\mathrm{km}$, focusing on the
properties of the postmerger remnant. Note that unlike a full-GR scheme,
the xCFC scheme does not require a large domain to apply outgoing
radiative boundary conditions or to ensure sufficient damping of the
constraint violation. Additionally, each refinement level in \BHAC has a
spherical structure that more accurately conforms to the shape of the
merger remnant than a Cartesian box used in \FIL.

\begin{figure}
  \includegraphics[width=0.49\textwidth]{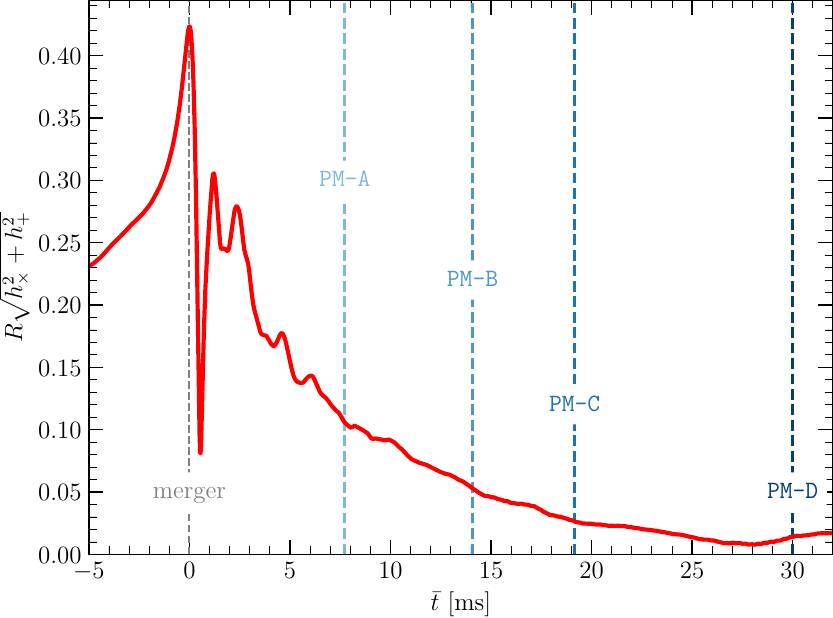}
  \caption{Evolution of the GW amplitude. Indicated with different dashed
    lines are either the time of merger or the different times when the
    \HO is performed.}
  \label{fig:gw_ho}
\end{figure}

\section{Validation of the \HO and RR}
\label{sec:results}

The CFC approach has been successfully applied in the study of uniformly
and differentially rotating neutron stars~\cite{Cook1996, Iosif2014,
  Iosif2020, Cheong2024, Cheong2024c}, with only minor deviations
observed compared to the full-GR solutions. Furthermore, our previous
work~\cite{Ng2024b} has demonstrated that the CFC approach can closely
approximate the full-GR solution in the highly nonlinear system of
neutron stars head-on collisions in full 3D simulations. However, the
performance of the CFC in realistic BNS postmerger scenarios with
magnetic fields and its comparison with full-GR simulations, has not been
explored thus far.

In this Section, we compare the evolution after \HO in \BHAC with the
corresponding evolution of the postmerger remnant in \FIL for four
different \HO times. In particular, after defining the time relative to
the merger as $\bar{t} := t - t_{\rm mer}$, we denote the time of \HO as
$\bar{t}_{_{\rm HO}}$ and consider the transfer of data from \FIL to
\BHAC at four different values of $\bar{t}_{_{\rm HO}}$, \ie from
\texttt{PM-A} to \texttt{PM-D} (this information is also summarised in
Tab.~\ref{tab:ID} and can be appreciated in
Fig.~\ref{fig:gw_ho}). Furthermore, to make the \HO times as robust and
reproducible as possible, we set them on the basis of the strength of the
GW radiation emitted by the postmerger remnant. More specifically, we \HO
the simulation from \FIL to \BHAC when the amplitude of the emitted GWs
decreases to $1/4$ (\texttt{PM-A}), $1/8$ (\texttt{PM-B}), $1/16$
(\texttt{PM-C}), and $1/32$ (\texttt{PM-D}) with respect to its peak
value (see also see Tab.~\ref{tab:global_diff}). Hereafter, special
attention will be paid on the results of the \HO made at the ``late''
time of $\bar{t} = 30 \, \rm ms$, \ie \texttt{PM-D}.

\subsection{On the role of \HO time}
\label{sec:diff_post_merger_time}

\begin{figure}
  \centering
  \includegraphics[width=0.48\textwidth]{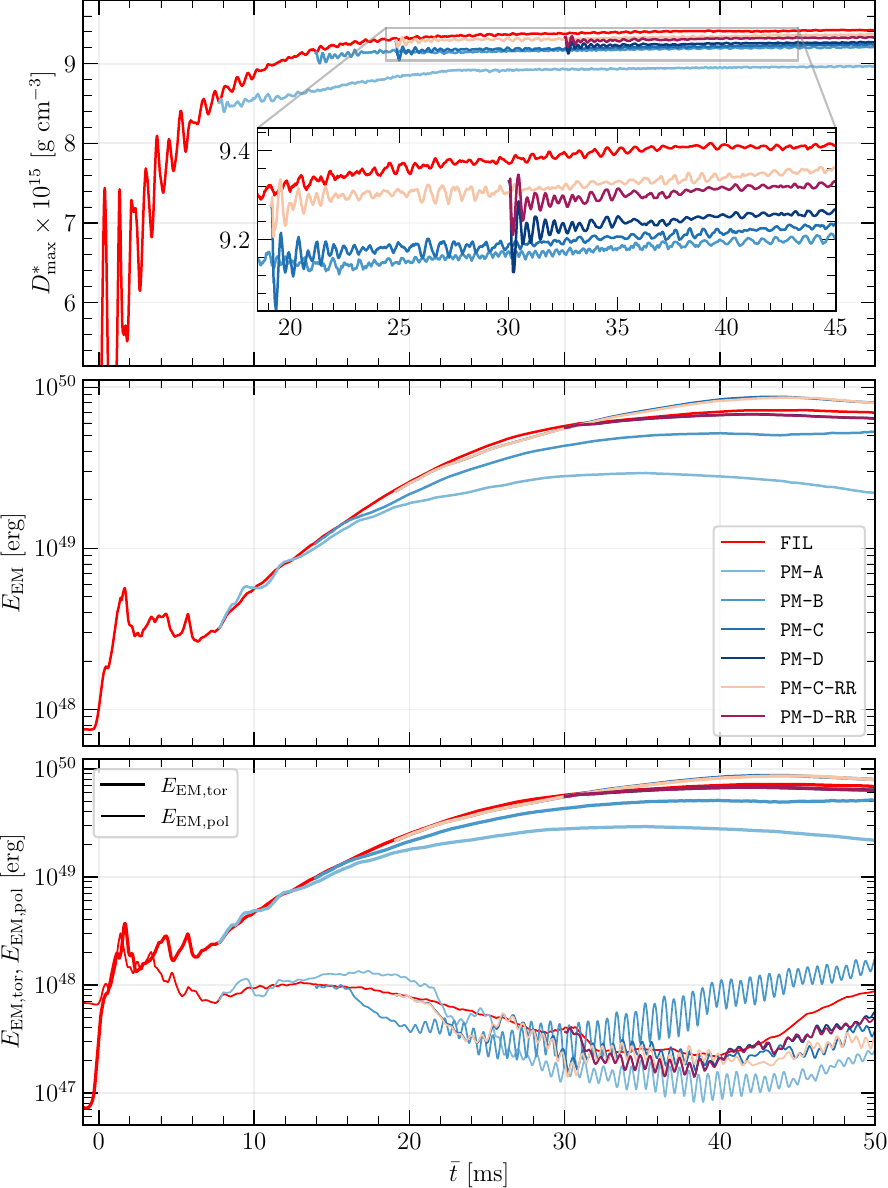}
  \caption{Evolution of the maximum value of the conserved rest-mass
    density $D^*$ (upper panel), of the total EM energy (middle panel),
    and of the toroidal/poloidal EM energies (bottom panel). Different
    lines refer to the values computed by \FIL (red solid line) and those
    by \BHAC at different \HO time cases, with and without RR
    (blue-shaded solid lines). Note that the evolutions \texttt{PM-C} and
    \texttt{PM-D} nearly overlap with the corresponding evolutions that
    include RR terms.}
  \label{fig:central_dense_post}
\end{figure}

In Fig.~\ref{fig:central_dense_post}, we report the evolution of three
representative fluid quantities as simulated by \BHAC with varying
$\bar{t}_{_{\rm HO}}$ values and compare them with the corresponding
results of \FIL, both when the RR terms are included or ignored. Starting
from the top panel of Fig.~\ref{fig:central_dense_post}, which reports
the maximum gauge-invariant conserved density $D^*_{\rm max}$, it is
possible to note that high-frequency oscillations are introduced
immediately following \HO, though they are quickly smoothed out by
subsequent evolution within $5~\mathrm{ms}$. The perturbation amplitudes
remain comparable across all \BHAC cases, suggesting that they primarily
originate from mapping errors rather than from gauge differences.

The evolution in \FIL reports a well-known dynamics, namely, that the
merger remnant reaches a quasi-equilibrium state after emitting energy
and transferring angular momentum, primarily through GW emission and
ejecta, and ultimately settling as a differentially rotating massive star
with a gradually increasing central density~(see~\cite{Kastaun2014,
  Hanauske2016, Iosif2021, Cassing2023} for a discussion of the
differential rotation profiles). At the earliest \HO time (\texttt{PM-A}
with $\bar{t}_{_{\rm HO}} = 7.5~\mathrm{ms}$), a rapid contraction occurs
in the remnant because the GW amplitude only decrease to $1/4$ of its
peak value. In this case, the use of xCFC without RR introduces a notable
relative difference of approximately $4.9\%$ in $D^*_ {\rm max}$ as it
reaches a quasi-equilibrium state at $\bar{t} =
25.2~\mathrm{ms}$. However, as the GW emission from the merger remnant
decreases, at $\bar{t} = 40 \,\mathrm{ms}$, the results from \BHAC align
well with \FIL even in cases without RR, with relative differences of
$2.1\%$, $1.6\%$, and $1.4\%$ in $D^*_{\rm max}$ for the cases
\texttt{PM-B}, \texttt{PM-C}, and \texttt{PM-D}, respectively.

Although all evolutions exhibit a gradual increase in $D^*_{\rm max}$,
those cases that do not include RR generally show slightly lower values
simply because the energy of the remnant does not have the possibility of
leaving the system and hence of reaching higher compactnesses, as noted
in~\cite{Ng2024b}. At the same time, the inclusion of RR terms, as for
cases \texttt{PM-C-RR} and \texttt{PM-D-RR}, does improve the expected
behaviour and leads to evolutions with a gradual increase in the maximum
rest-mass density starting from right after the \HO and ultimately
reaching a quasi-equilibrium state similar to \FIL, with relative
differences of $0.32\%$ and $0.74\%$ in $D^*_{\rm max}$ at $\bar{t} =
40~\mathrm{ms}$, respectively. Overall, these simulations demonstrate
that the inclusion of RR terms is important to reach a postmerger
evolution that is as faithful as possible.

The middle panel of Fig.~\ref{fig:central_dense_post} reports the
evolution of the EM energy [see definition in Eq.~\eqref{eq:em_total}]
and illustrates that the magnetic-winding process is very well captured
by \BHAC in cases when the \HO is done when the GW amplitude decreases
below $1/16$ of its peak value, \ie cases \texttt{PM-C} and
\texttt{PM-D}. This accurately represents both the rate of change in EM
energy and the time at which winding ceases. Indeed, the largest relative
differences in total EM energy is of only $4.7\%$ for \texttt{PM-D}, but
obviously increases to $15.4\%$ and $31.0\%$ when the \HO is done
earlier, \ie for cases \texttt {PM-C} and \texttt{PM-B}, respectively.

Finally, in the bottom panel of Fig.~\ref{fig:central_dense_post} we
report the evolution of the toroidal and poloidal EM energies [see
  definition in Eq.~\eqref{eq:em_tor}]. As well-known, the toroidal
magnetic field dominates the postmerger magnetic energy as a result of
the KHI and the winding process, which mostly convert kinetic energy into
toroidal magnetic energy. On the other hand, the poloidal EM energy
continues to dissipate until the winding process ends, due to a
combination of shocks and violent oscillations within the remnant, with a
mild increase at later times. Interestingly, oscillations in the
evolution of poloidal EM energy emerge in \BHAC if the \HO is performed
too early. These oscillations reflect the development of an $\ell=2, m=1$
barmode instability~\cite{East2015, Lehner2016, Radice2016a, East2019,
  Papenfort:2022ywx, Topolski2024} triggered by the error introduced at
\HO if this is made too early. Indeed, the amplitude of the oscillations
is reduced for later hand-offs (the smallest amplitude is observed in
\texttt{PM-D}, due to greater axisymmetry and conformal flatness (see 
also a discussion in Appendix~\ref{sec:cy}) at this stage) and it decays
rapidly with time in all cases. More importantly, because these
oscillations only affect the poloidal EM energy and the latter is about
two orders of magnitude smaller than the toroidal one, the overall
evolution is hardly affected (see middle panel).

In summary, our results show that using the GW amplitude represents
a gauge-invariant, model-independent, and essentially
resolution-independent manner to assess when the \HO can be
performed. In addition, we suggest that performing the \HO when the GW
amplitude has reached $\simeq 1/16$ of the peak amplitude represents a
robust choice. While these conclusions are based on a limited number of
tests, the underlying motivation is clear: the smaller the GW emission,
the better the CFC approximation and its use at \HO. At the same time,
we also recommend that some experimentation is made around the \HO
time. While this experimentation brings in some extra computational
costs, the latter are minute when compared with those associated with
the long-term evolution that can be performed under optimal \HO
conditions.

\subsection{Case \texttt{PM-D}: xCFC vs full-GR results}
\label{sec:detail_PostD}

\begin{figure}
  \centering
  \includegraphics[width=0.48\textwidth]{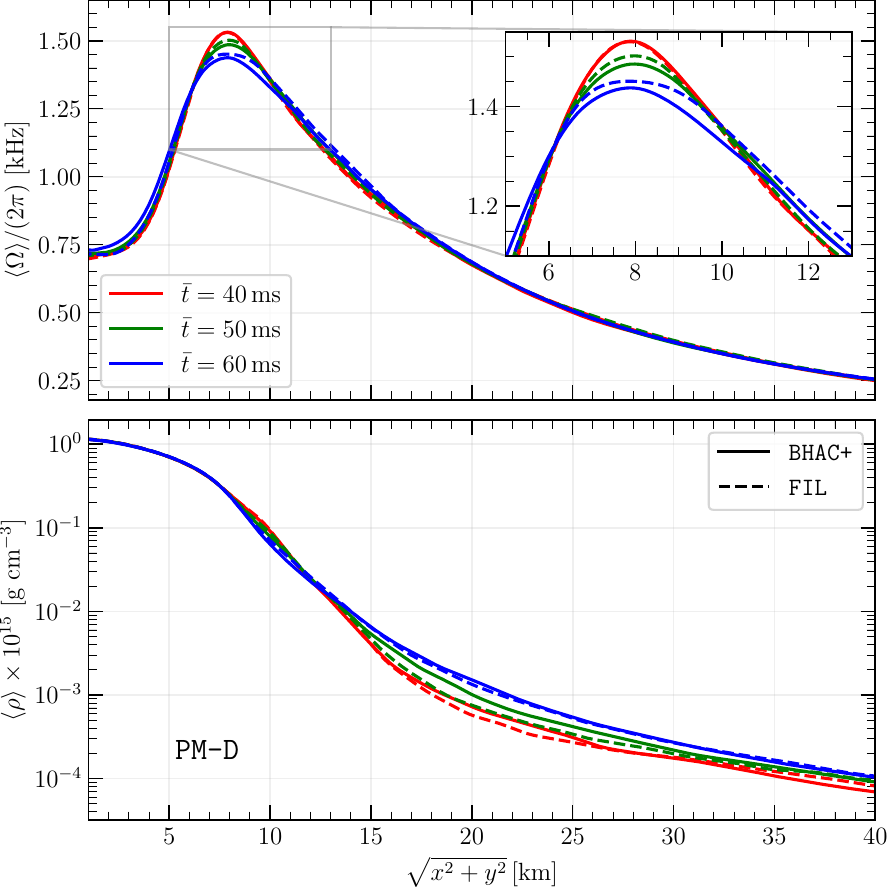}
  \caption{Azimuthally averaged profiles of the angular velocity (upper
    panel) and of the rest-mass density (lower panel) as computed at
    different times by \FIL (dashed lines) and by \BHAC (solid
    lines). The data is computed on the equatorial plane and refers to
    the case \texttt{PM-D}, \ie with a \HO at $\bar{t}_{_{\rm HO}}=30 \,
    \rm ms$.}
  \label{fig:phi_ave_profile_post}
\end{figure}

As discussed above, the \HO at $\bar{t}_{_{\rm HO}}=30\, \rm ms$ provides
the best match with the full-GR evolution computed by \FIL, thus making
it worth a closer investigation. Hence, we start our comparison between
the results of \BHAC (solid lines) and \FIL (dashed lines) by presenting
in Fig.~\ref{fig:phi_ave_profile_post} the azimuthally averaged profiles
of the rest-mass density and of the angular velocity on the equatorial
plane of the merger remnant at different times after \HO. The
differential-rotation profiled observed (top part of
Fig.~\ref{fig:phi_ave_profile_post}) is well known~\cite{Kastaun2014,
  Hanauske2016, Cassing2023} and shows a peak at $r \approx 8\,\rm{km}$,
a Keplerian fall-off of the type $r^{-3/2}$, and gradually flattens over
time due to winding and dissipation in both codes~\cite{Duez2004b,
  Duez:2005cj, Ciolfi2019} (see also the discussion in
Sec.~\ref{sec:flattening}). A measure of the very good match between the
two evolutions can be found in the relative difference in the rotation
frequency, which remains below $2\%$ ($5\%$) at $20\, \rm ms$ ($30\, \rm
ms$) after the \HO. Similarly, the difference in the location of the
angular-velocity maximum between the two codes remains within $1\%$
across all the times considered, indicating that angular momentum
redistribution and frame-dragging effects are similarly captured in both
codes at this stage.

The different profiles of the rest-mass density produced by the two codes
(bottom part of Fig.~\ref{fig:phi_ave_profile_post}) are essentially
identical in the core regions within $r \approx 8\,\rm{km}$ and do not
vary significantly at least within the first $30\,{\rm ms}$ following the
\HO. In the outer regions of the remnant, matter expands outward through
oscillations, leading to an increase in rest-mass density over time for
both codes. The largest relative difference remains below $1.5\%$ within
$r \approx 8\, \rm km$ (where $96.4\%$ of the rest-mass is concentrated)
and still remains below $10\%$ within $r \approx 15\, \rm km$ (where
$97.2\%$ of the rest-mass is concentrated). Larger differences appear at
large radii but these are related to the slightly different locations of
the spiral shocks and involve amounts of matter that are not dynamically
relevant.

\begin{figure*}
\center
\includegraphics[width=0.49\textwidth]{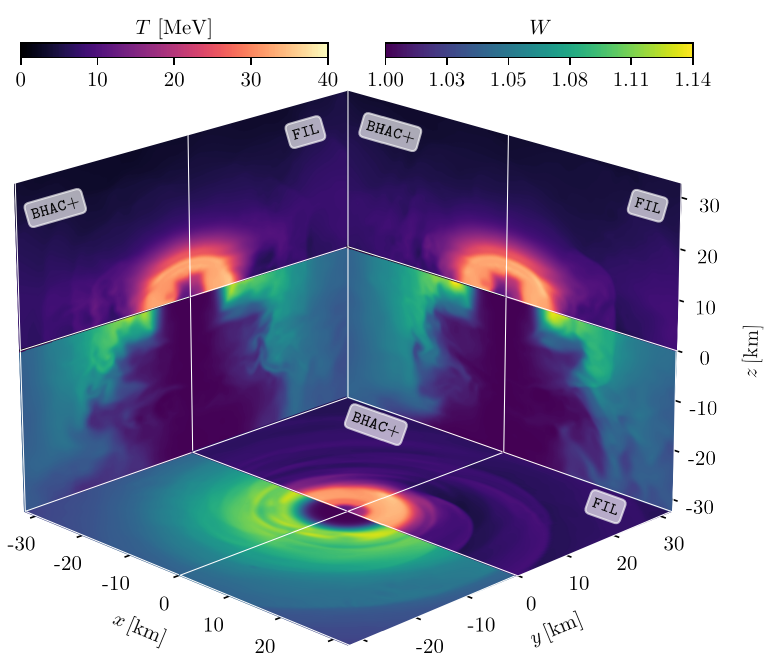}
\hskip  0.25cm
\includegraphics[width=0.49\textwidth]{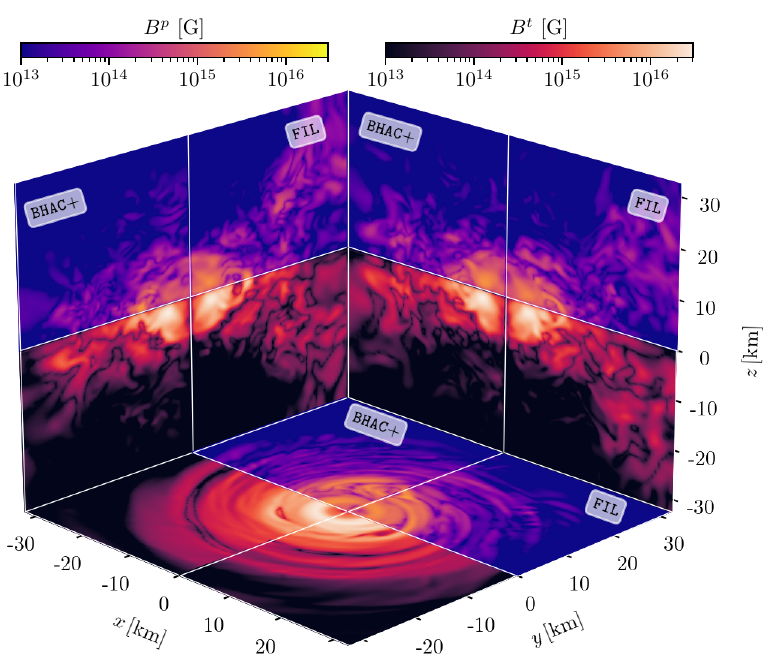}
\caption{Projections on the three principal planes of the temperature and
  Lorentz factor (left panel) and of the toroidal and poloidal
  magnetic-field strengths (right panel) for the merger remnant from case
  \texttt{PM-D} at $20\, \rm ms$ after \HO. For each plane, the left
  portion refers to simulations from \BHAC and the right one from \FIL.
  Note that the negative-$z$ regions are included for visualization
  purposes only, as the simulations assume symmetry across the $z=0$
  plane.}
\label{fig:post_comparison_3d}
\end{figure*}

Figure~\ref{fig:post_comparison_3d} provides an additional visual
impression of the very good match between the two codes by showing 2D
projections onto the principal planes of the temperature $T$, the
Lorentz factor $W$ (left panel), and magnetic-field strengths in the
poloidal and toroidal components (right panel) at $20\, \rm ms$ after
\HO. When comparing the temperature distributions (left panel of
Fig.~\ref{fig:post_comparison_3d}), the well-known ``hot shell'' is
formed in both cases and gradually expands both inward and outward as the
evolution progresses due to turbulent mixing and shock heating. We
observe overall agreement, although some slight differences are
noted. First, there is a temperature difference of $\sim 2\, \rm MeV$ for
the hot shell structure, and a larger difference of $\sim 4\, \rm MeV$ in
the central region where $\rho \gtrsim3~\, \rho_{\rm sat}$ is reached.
Additionally, the hot shell structure in \BHAC is $\sim 0.8\, \rm km$
larger than that in \FIL. Given the different numerical methods and set of
equations solved, it is remarkable that the differences between the two
codes on a derived, \ie non-primarily evolved quantity, remain so small.
As for the Lorentz factor $W$, the profiles once again match well between
the two codes, as expected from the well-matching rotation profiles. The
highest speed region forms a ring structure on the equatorial plane with
the highest Lorentz factor $W \simeq 1.14$, while on the latitudinal
planes, the highest speed region forms a wing-shaped structure. The
funnel region has an opening angle of $\sim 60^{\circ}$ and does not yet
exhibit a high-speed outflow in either code. Additionally, we observe
that the Lorentz factor's ring structure ($6.3$--$10.8\, \rm km$) lies
moderately outside the temperature ring structure ($4.0$--$7.4\, \rm
km$), indicating strong shear in the hottest regions.

Similar considerations can be made also for the distributions of the poloidal
and toroidal magnetic field (right panel of
Fig.~\ref{fig:post_comparison_3d}). In this case, the most significant
feature is a strong toroidal magnetic field forming a ring structure
located between $\sim 3.4\, \rm km$ and $\sim 6.2\, \rm km$ due to
winding caused by strong differential rotation. This prominent structure
is reproduced in both codes and the match between the two is very good,
despite the use of different methods for evolving the magnetic
field~\cite{Etienne2015, Most2019b, Olivares2019}. In the low-density
regions (\ie $\rho \ll \rho_{\rm sat}$), the magnetic field is dominated
by turbulence and we do not expect the distributions of magnetic field in
the two codes to have a significant resemblance. Yet, the statistical
properties of the turbulent magnetic field in the two codes is very
similar, as can be easily appreciated in the right panel of
Fig.~\ref{fig:post_comparison_3d}. We also note that, as mentioned
above, the toroidal magnetic-field strength is systematically stronger
than that of the poloidal magnetic field on all three principal planes
(see also the bottom panel of Fig.~\ref{fig:central_dense_post}). For
example, on the equatorial plane, the maximum values are $2.6\times
10^{16}\, \rm G$ and $3.2\times 10^{15}\, \rm G$ for the toroidal and
poloidal components, respectively. As discussed in several works in the
literature~\cite{Ciolfi2019, Palenzuela_2022PRD, Chabanov2022,
  Kiuchi2024, Musolino2024b}, this difference between the poloidal and
toroidal magnetic fields arises for two reasons: first, the winding
process significantly amplifies the toroidal magnetic field; second, the
lack of an efficient dynamo process that limits the conversion of
toroidal to poloidal magnetic field in the current simulations.

In summary, the results discussed in this Section provide very strong
evidence that \BHAC, despite a simpler description of the gravity sector,
namely, the use of the CFC approximation reproduces the full-GR results
coming from \FIL extremely well in all fluid quantities. In addition, the
inclusion of RR terms in the xCFC scheme further improves the accuracy.
Furthermore, while these considerations have been deduced when
considering the ``late'' \HO at $\bar{t}_{_{\rm HO}}=30 \, \rm ms$,
\texttt{PM-D} (when the GW amplitude has reached $\sim 3\%$ of its peak
value), similar considerations can be made also with earlier hand-offs
(see Appendix~\ref{sec:early_ho} for a similar discussion in the case of
the \HO \texttt{PM-C}). 

As a final remark, we note that the differences reported here in the
postmerger evolutions from \FIL and \BHAC are significantly smaller than
those reported in Refs.~\cite{Espino2023, Neuweiler2024} in terms of the
evolution of the maximum rest-mass density, total EM energy,
magnetic-field strength, rotational profiles, and remnant lifetimes among
different GRMHD codes. Despite all of these codes employed in
Refs.~\cite{Espino2023, Neuweiler2024} make use of full-GR treatments and
adopt similar numerical schemes, differences remain and underline the
difficulties in providing very accurate descriptions of the postmerger
evolution.

\section{Impact of magnetic-field strength on long-lived ``magnetars''}
\label{sec:long_term}

While the previous Sections have been dedicated to the discussion and
testing of the strategy developed to optimise the hybrid strategy
proposed in Ref.~\cite{Ng2024b} to perform long-term simulations of
postmerger remnants, the remainder of the paper is dedicated to use the
\FIL--\BHAC infrastructure to actually explore a specific issue on the
evolution of the postmerger remnant, namely, the impact of the
magnetic-field strength on the dynamics and EM emission from a long-lived
highly magnetized merger remnant, or ``magnetar'', produced in a BNS
merger.

We recall that the magnetic-field strengths in the BNS merger remnant are
expected to reach up to equipartition values of $\sim
10^{16}$-$10^{17}\,\rm G$, given the large energy reservoir in the
remnant, both in terms of kinetic and binding energy, and that current
resolutions of numerical simulations still can only partially resolve two
key amplifying effects, i.e., KHI and MRI, of the magnetic field. This is
shown by sub-grid modelling simulations
~\cite{Aguilera-Miret2021,Palenzuela_2022PRD}, or simulations adding
phenomenological dynamo terms~\cite{Shibata2021c, Most2023, Most2023b},
and also very high-resolution simulations~\cite{Kiuchi2015a,
  Chabanov2022, Kiuchi2023, Musolino2024b}. Given these limitations and
the enormous costs of carrying out simulations with very high resolution
and on very long timescales, we here consider a different
approach. Namely, we consider and contrast two different scenarios that
differ only in the strength of the initial magnetic field, i.e.,
$2.0\times10^{16}\, \rm G$ and $1.0\times10^ {17}\, \rm G$, so as to
assess what is the role of the amplification mechanism -- whatever that
is -- on the subsequent long-term evolution of the postmerger. Hereafter,
we will refer to these two initial magnetic fields as low-magnetic-field
(LMF) and high-magnetic-field (HMF), respectively.

More specifically, for each initial magnetic-field strength, we have
performed a simulation with \FIL with the finest resolution of $300\, \rm
m$ of the inspiral, merger and postmerger. Such a solution was then
transferred to \BHAC with a \HO at $20\, \rm ms$ (this is when the GW
amplitude has dropped below $1/16$ of its peak value) and then evolved
with two different resolutions of $300\, \rm m$ and $150\, \rm m$ so as
to have a postmerger evolution over a timescale of $200\,{\rm ms}$ with a
resolution that would be prohibitive with \FIL. More specifically, in
high-resolution simulations, we add the finest refinement level in the
region $[-16,\, 16]\,\rm km$, which adequately covers the region where
$\rho > 10^{12} \, \rm g \, cm^{-3}$. Hereafter, we will refer to these
four scenarios respectively as \texttt{LowB-LR}, \texttt{LowB-HR},
\texttt{HighB-LR}, and \texttt{HighB-HR} (see also Tab.~\ref{tab:ID}).

\subsection{Impact on the EM-energy evolution}

\begin{figure}
  \center
  \includegraphics[width=1.0\columnwidth]{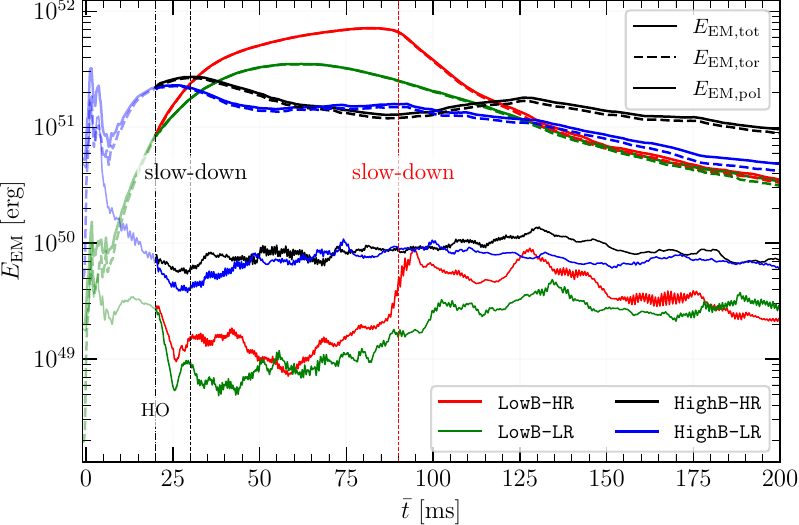}
  \caption{EM-energy evolution for simulations with LMF and
    HMF. Different line-styles represent different classes of EM energy,
    namely, $E_{\rm EM,tot}$ (thick solid line), $E_{\rm EM,tor}$ (thick
    dashed line), and $E_{\rm EM,pol}$ (thin solid line), while
    different colours indicate simulations that differ either in
    resolution or initial magnetic field (see Tab.~\ref{tab:ID} for the
    nomenclature of the various cases). The black vertical dash-dotted
    line marks the \HO time, with the results from \BHAC (\FIL) shown on
    the right (left) with respect to it. The vertical black (red) dashed
    line indicates the time at which the remnant undergoes a slow down at
    the peak of EM energy in the HMF (LMF) scenario. }
  \label{fig:em_evolve}
\end{figure}

We start our discussion by reporting in Fig.~\ref{fig:em_evolve} the
evolution of the EM energy in the four merger remnants. Lines of
different colours refer to the four scenarios considered and the black
vertical dash-dotted line represents the \HO time, with the results from
\BHAC (\FIL) shown on the right (left) with respect to it. Note that, as
remarked above, the \texttt{LowB-LR} and \texttt{HighB-LR} simulations in
\FIL are then also evolved at higher resolution in \BHAC as
\texttt{LowB-HR} and \texttt{HighB-HR}; this explains why there are only
two sets of lines to the left of the \HO time.

Both the toroidal and poloidal EM energy grow rapidly during the KHI
stage, peaking at around $\bar {t} \approx 2\, \rm ms$. After this point,
the EM energy begins to rapidly decrease, mostly due to the lack of
sufficient turbulent vortices to counteract dissipation (see discussion
in~\cite{Chabanov2022}). As the merger remnant develops a large-scale
differential rotation structure, the winding stage commences and the
toroidal magnetic-field strength exhibits the characteristic linear
growth over time, which can last up to $\bar{t}\simeq 30\,{\rm ms}$ ($\bar{t}\simeq
90\,{\rm ms}$) for the HMF (LMF) case at high resolution. Over the same
period of time, we observe a rapid decay in the poloidal-energy
component, primarily due to numerical dissipation caused by a combination
of shocks and violent oscillations within the remnant.

Magnetic winding ceases when the toroidal magnetic fields become strong
enough to exert an effective torque on the fluid, resulting in angular
momentum redistribution. This transition, which we denominate as
``slow-down'' occurs earlier in the HMF case (\ie $\bar{t}\simeq 30\,{\rm ms}$)
than in the LMF case ($\bar{t}\simeq 90\,{\rm ms}$); the relevant times are
marked with vertical dashed lines in Fig.~\ref{fig:em_evolve}. It is
important to underline that the slow-down does not mark the end of the
magnetic-field winding, which indeed continues also afterwards, but
rather the beginning of a different regime in which winding of the
poloidal magnetic field into the toroidal one still takes place but is
slowed down, \ie it leads to a sublinear in time growth, because of the
intense back-reaction of the magnetic tension (see also the discussion in
Sect.~\ref{sec:flattening}). In addition, we should note that we expect
the specific threshold depends, in addition to the initial magnetic-field
strength, also on the rotational profile of the merger remnants and hence
also on the mass ratio and EOS. This redistribution of angular momentum
simultaneously generates an effective poloidal fluid motion,
counteracting the dissipation of poloidal EM energy. Additionally, there
is another contribution to the poloidal EM-energy growth after the EM
amplification ceases and it comes from buoyant toroidal field lines
that rise to higher latitudes, causing magnetic flares that build
structured poloidal magnetic fields in the polar region (see discussion
below).

An important and interesting behaviour found is that when the initial
magnetic field is not large, \ie in the LMF case, the winding period
lasts longer and achieves higher EM energy than the HMF
case. Specifically, the high-resolution simulations show peak EM energies
of $2.7 \times 10^{51}\, \rm erg$ for the HMF case and $7.1 \times
10^{51}\, \rm erg$ for the LMF case, hence with an EM energy that is
about three times larger. Naively, one would expect that starting with a
weaker magnetic field would also yield a lower final magnetic field,
which is the opposite of what our simulations have revealed. The
explanation of this counter-intuitive behaviour has to do with the
nonlinearity of the amplification process and on the ability to use the
available kinetic energy stored in the fluid to convert it into magnetic
energy. In particular, these results show that when the initial magnetic
strength is large, it exerts a substantial feedback on fluid motion,
suppressing small-scale turbulence through magnetic tension, thus
preventing winding and effectively weakening the amplification very
rapidly or, equivalently, leading to an early slow-down (the simple
linear winding essentially stops around $\bar {t} \approx 30\, \rm ms$
for the HMF case). On the contrary, when the initial magnetic strength is
small, the fluid is less constrained in its dynamics and can be used for
a larger number of winding periods, thus leading to a stronger final
magnetic field and a delayed weakening of the amplification (the linear
winding essentially stops around $\bar {t} \approx 90\, \rm ms$ for the
LMF case). This interpretation of the dynamics is also supported by the
analysis of the fraction of the kinetic energy that has been converted
into EM energy, which is only $\simeq 2\%$ in the HMF case and $\simeq
6\%$ in the LMF case. Finally, we note that the use of higher resolution
has the effect of prolonging the winding period, delaying the appearance
of the slow-down time; this difference is of $\sim 6\,\rm ms$ or $20\,\rm
ms$ in the HMF case and LMF cases, respectively. This extension of the
winding can be explained by the fact that higher resolution naturally
reduces numerical resistivity, thus allowing for more efficient bending
of the magnetic field lines via differential rotation and hence to a more
efficient winding (see also the discussion in
Refs.~\cite{Dionysopoulou2015, Shibata2021c}).

\begin{figure}
 \center
 \includegraphics[width=1.0\columnwidth]{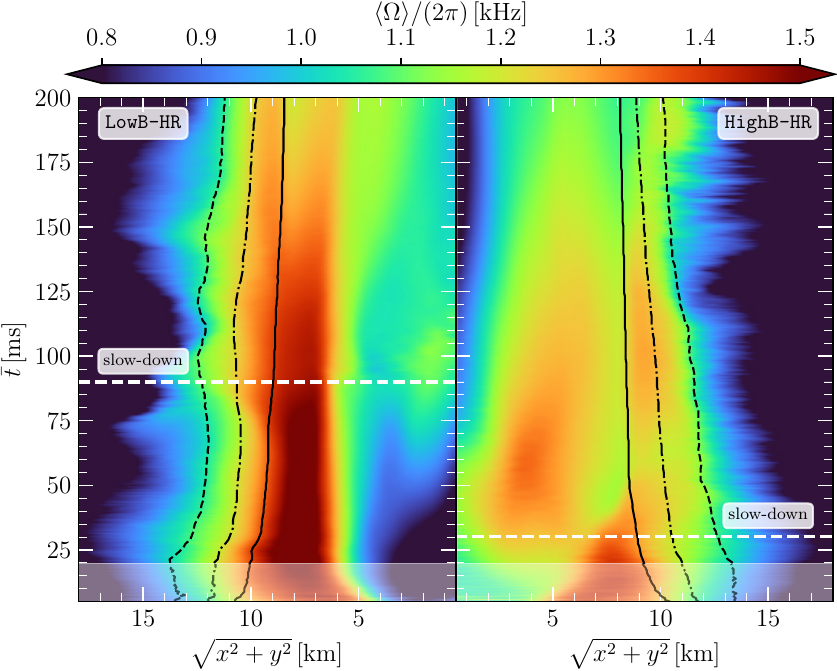}
 \caption{Spacetime diagram of azimuthally averaged sections of the
   angular-velocity profile on the equatorial plane of the remnant for
   the LMF (left panel) and the HMF scenarios (right panel). The
   transparent region reports the result of \FIL, while the rest of the
   diagram reports \BHAC data. The horizontal dashed white lines
   represent the time of slow-down, while the dashed, dot-dashed, and
   solid black lines report the worldlines of rest-mass density contours
   at $10^{13}, 10^{13.5}, \text{and}\, 10^{14}\, \rm g~cm^{-3}$,
   respectively.}
 \label{fig:rot_profile}
\end{figure}

\subsection{Impact on the differential-rotation evolution}
\label{sec:flattening}

It is well-known that the interaction between differential rotation and
magnetic fields in stars inevitably leads to the generation of magnetic
field, but also to the suppression of differential rotation (see
Refs.~\cite{Rezzolla00, Shapiro00, Rezzolla01a, Duez:2005cj}, and also
Refs.~\cite{Duez2004b, Radice2017} for an equivalent process in the
presence of shear viscosity). It is therefore interesting to investigate
how the process of removal of differential rotation takes place in a BNS
remnant and how this depends on the (initial) strength of the magnetic
field.

To this scope, we report in Fig.~\ref{fig:rot_profile} a spacetime
diagram over the whole timescale of the simulation of azimuthally
averaged sections of the angular-velocity profile on the equatorial plane
of the remnant for the LMF (left panel) and the HMF scenarios (right
panel), respectively. Reported in the white-shaded area are the results
from \FIL, while the data after \HO refers to the evolution obtained with
\BHAC. In such a spacetime diagram it is particularly interesting to
contrast the early part of the evolution at $\bar{t} \sim 25\,{\rm ms}$
with the final ones at $\bar{t} \sim 200\,{\rm ms}$. Clearly, the remnant
has a significant amount of differential rotation at first (see also
Fig.~\ref{fig:phi_ave_profile_post}) but this is gradually changed by the
interaction between the rotation of the fluid and the magnetic-field
tension. As a result, in both scenarios of low and high initial magnetic
field, the degree of differential rotation is significantly smaller and
differential rotation is essentially removed in the HMF case by the time
the simulation is ended.

\begin{figure*}
  \center
  \includegraphics[width=0.43\textwidth]{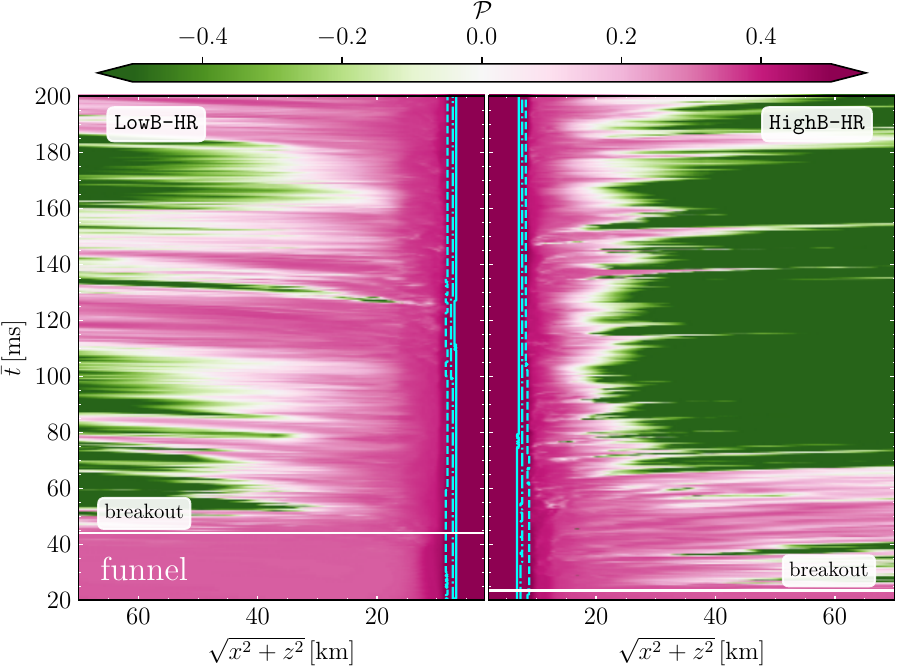}
  \hskip  0.25cm
  \includegraphics[width=0.43\textwidth]{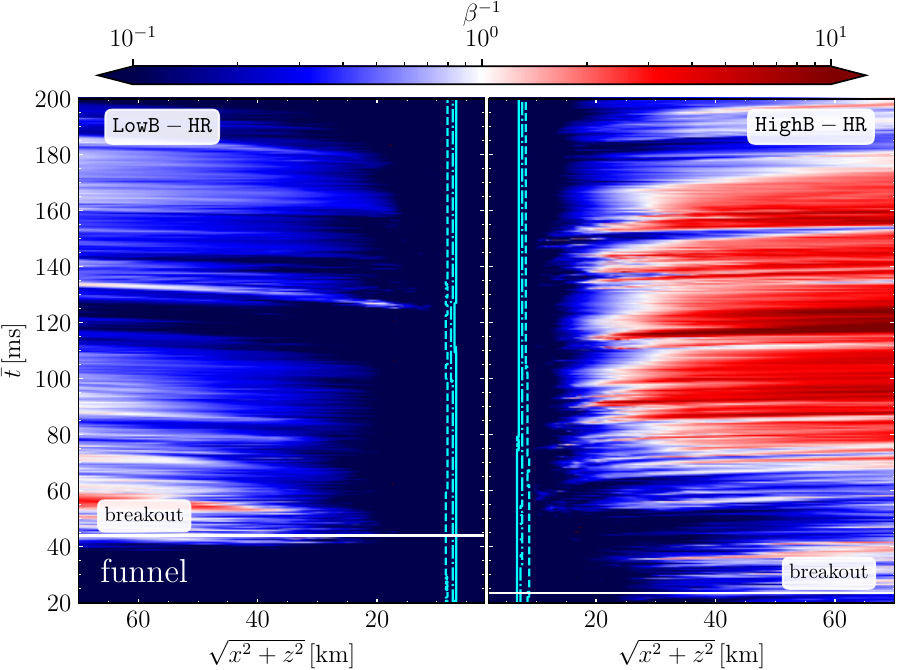}
  \vskip  0.25cm
  \includegraphics[width=0.43\textwidth]{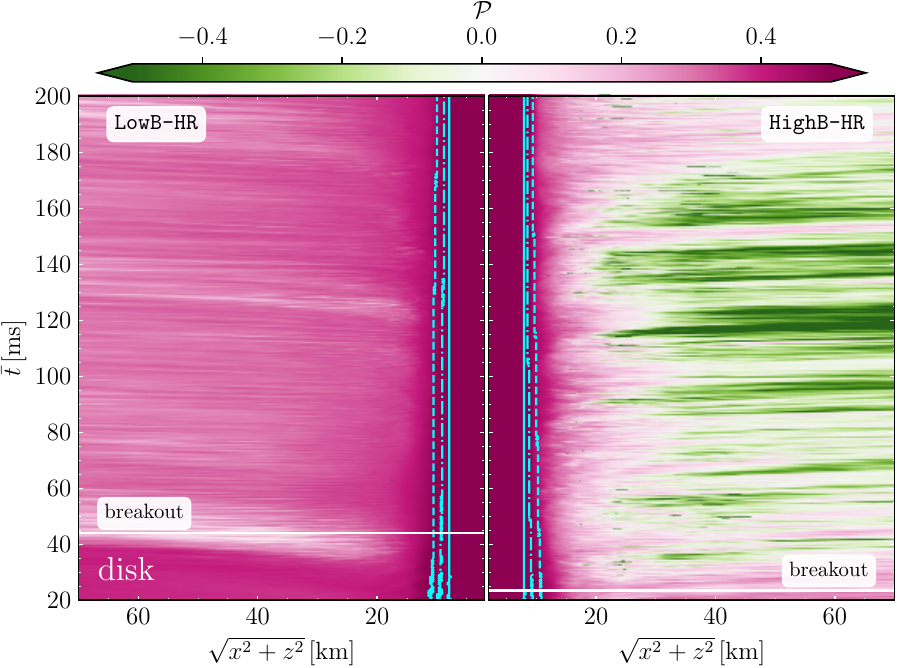}
  \hskip  0.25cm
  \includegraphics[width=0.43\textwidth]{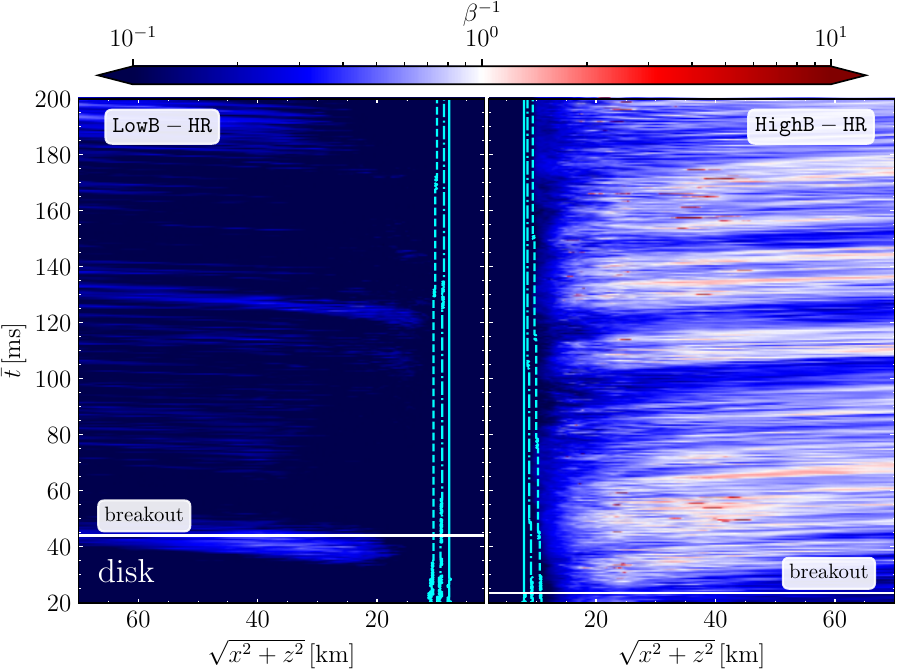}
  \caption{Spacetime diagrams of the Parker-instability criterion (left
    panel) and of the inverse plasma-$\beta$ (right panel) averaged on
    the $(x,z)$ plane in simulations with initial LMF (left portions) and
    HMF (right portions), respectively. The top part of the figure refers
    to the ``funnel'' region, \ie the region with a polar angle less than
    $\pi/6$, while the bottom part refers to the ``disk'', which is the
    remainder. The breakout times are shown with horizontal white solid
    lines for each simulation. Vertical dashed, dash-dotted, and solid
    cyan lines represent the worldlines of rest-mass density contours at
    $10^{13}, 10^{13.5}, \text{and}\, 10^{14}\, \rm g~cm^{-3}$,
    respectively.}
  \label{fig:parker_c}
\end{figure*}

Note also the strong correlation that we find between the onset of the
slow-down (\ie the quenching of the linear-in-time winding discussed
above) and the transition to process of removal of differential
rotation. This is not surprising, since the slow-down marks the time when
linear-in-time winding ceases because magnetic tension is exerting a
sizeable torque on the fluid and opposing the generation of additional
toroidal magnetic field. Another interesting aspect of the suppression of
differential rotation that can be appreciated from the spacetime diagram
in Fig.~\ref{fig:rot_profile} is that the remnant also undergoes a
significant redistribution of angular momentum, which can either increase
the angular velocity in the core (LMF case) or decrease it. These
structural changes in the angular momentum also show that the initial
peak in differential rotation is not simply washed out. Rather, before
the angular velocity attains an almost uniform rotation profile (see the
HMF case in Fig.~\ref{fig:rot_profile} at $\bar{t} \gtrsim 150\,{\rm
  ms}$) two different peaks appear at smaller and larger radii, which are
progressively erased (this can be seen also in the LMF case in
Fig.~\ref{fig:rot_profile} but takes place much later). Hence, the
``forking'' of the angular-velocity maximum can also be taken as a good
proxy of the loss of efficiency in the winding and the beginning of the
slow-down phase. In addition, the redistribution of angular momentum can
also be tracked via the worldlines of the rest-mass density at $10^{13}$
(black dashed line), $10^{13.5}$ (black dot-dashed line), and $10^{14}\,
\rm g~cm^{-3}$, which indicate that all of the corresponding iso-contour
are moving inwards as a result of the depletion of angular momentum in
the remnant and that is transported outwards as a result of the MRI and
winding.

Overall, our results are in good agreement with what has been observed in
simulations with added effective shear~\cite{Fujibayashi:2020dvr}, or
resistivity~\cite{Dionysopoulou2015, Shibata2021c}, and with very strong
magnetic fields~\cite{Bamber2024b}. When taken together, this bulk of
works suggests that differential rotation in a BNS remnant is likely to
be washed out over a timescale that clearly depends on the strength of
the magnetic field but that can be reasonable narrowed between $\bar{t}
\sim 200\,{\rm ms}$ for magnetic-fields of the order of $\simeq
10^{17}\,{\rm G}$ and $\bar{t} \sim 300\,{\rm ms}$ for weaker
magnetic-fields of $\simeq 10^{16}\,{\rm G}$. However, we caution that
when starting from realistic initial-field strengths of $\simeq
10^{11}\,{\rm G}$, the winding process might be prolonged due to the
turbulent magnetic field in the early postmerger phase.

\subsection{Parker instability and flares}
\label{sec:parker}

As mentioned in Ref.~\cite{Musolino2024b}, the physical conditions in the
outer layers of the merger remnant and close to the polar axis can lead
to a global breakout of the plasma as a result of the development of the
Parker instability. We recall that the latter is described by the
criterion~\cite{Parker1966}
\begin{equation}
  \label{eq:parker_c}
  \mathcal{P} := \frac{d\log{p}}{d\log{\rho}} - 1 - \frac{\beta^
   {-1}\left(1 + 2\beta^{-1}\right)}{2 + 3 \beta^{-1}}\,,
\end{equation}
where $\beta := p / (b^2 / 2)$ is the ratio of the fluid pressure to the
magnetic pressure. If the magnetic-field pressure continues to increase
within a region, the third term in Eq.~\eqref{eq:parker_c} becomes
dominant and once the condition for instability $\mathcal{P} < 0$ is met,
low-density matter with high total pressure will rise due to buoyancy,
carrying the magnetic field along with it. We refer to this process as
to the ``breakout'' as it characterises a sharp transition between two
different states of the plasma in the polar region.

\begin{figure*}
  \center
  \includegraphics[width=0.98\textwidth]{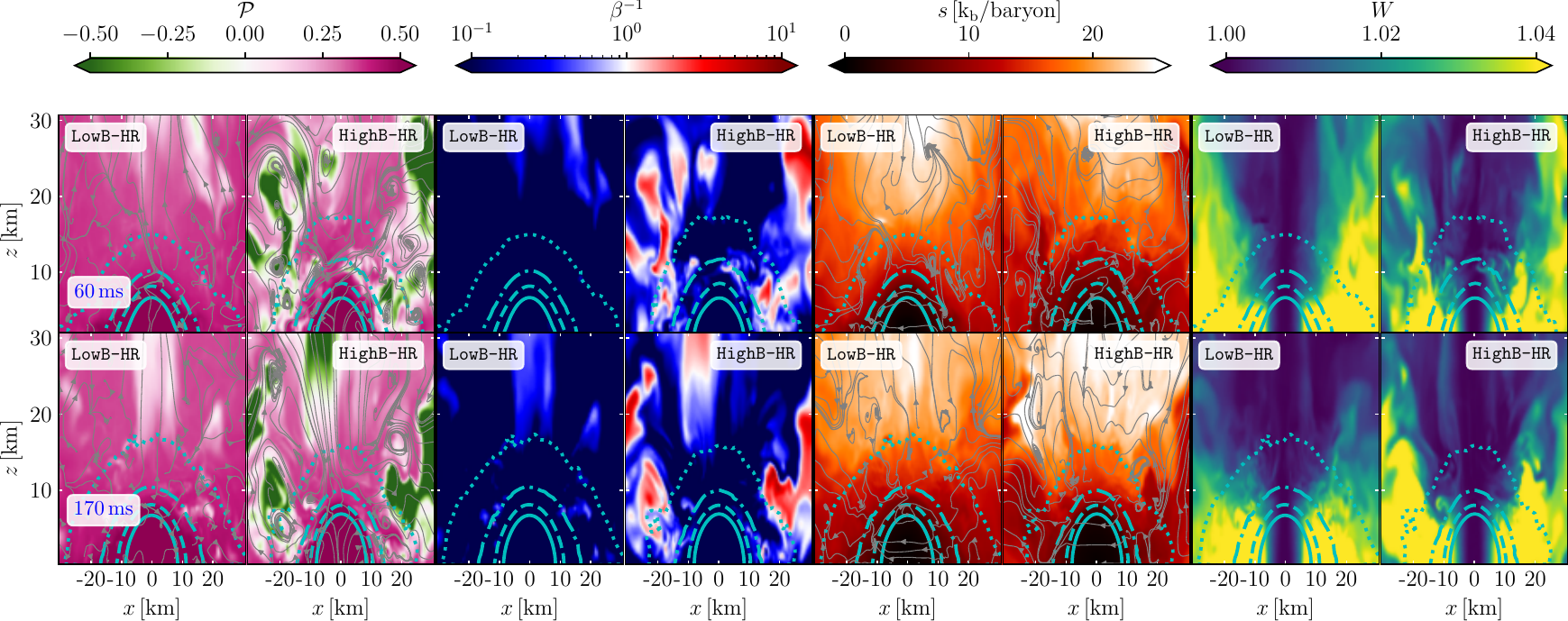}
  \caption{Magnified section on the $(x,z)$ plane of the merger remnant
    at $\bar{t}=60\, \rm ms$ (top row) and $\bar{t}=170\, \rm ms$ (bottom
    row) so as to highlight the generation of the magnetic eruptions
    (flares). Starting from the left, the different columns report: the
    Parker instability criterion $\mathcal{P}$, the inverse
    plasma-$\beta$, the specific entropy $s$, and the Lorentz factor $W$.
    Magnetic-field lines and fluidlines are shown in the first and third
    columns, respectively. For each subpanel, the dotted, dash-dotted,
    dashed, and solid cyan lines show the rest-mass density contour at
    $10^{11}, 10^{12}, 10^{13}, \text{and}\, 10^{14}\, \rm g~cm^{-3}$,
    respectively.}
  \label{fig:flare_view}
\end{figure*}

To analyze the occurrence of the Parker instability, we use the
approach proposed in Ref.~\cite{Musolino2024b} to divide the
simulation domain into two parts: the region with a polar angle less than
$\pi/6$, defined as the ``funnel'', and the remainder, defined as the
``disk''. Figure~\ref{fig:parker_c}, in particular, reports the spacetime
diagrams of the Parker-instability criterion (left panels) and of the
inverse plasma-$\beta$ (right panels) for simulations with different
magnetic-field strengths. Both quantities are averaged over the polar
angle on the $(x,z)$ plane. While the top plots refer to the funnel
region, the bottom ones show the disk region; at the same time,
for each plot, the left and right portion refers to the LMF and HMF case,
respectively. The breakout time, \ie the time when $\mathcal{P}$
changes sign, is shown with horizontal white solid lines for each
simulation, while vertical dashed, dash-dotted, and solid
cyan lines represent the worldlines of rest-mass density contours of
$10^{13}, 10^{13.5}, \text{and}\, 10^{14}\, \rm g~cm^{-3}$, respectively.

Using the spacetime diagrams, it is then straightforward to recognise
that a breakout clearly takes place in the funnel region (top panels in
Fig.~\ref{fig:parker_c}) and that this is true for the LMF and the HMF
scenarios. What varies between these two cases is mostly the time of the
breakout, which obviously takes place earlier for the HMF scenario
($\simeq 25\, \rm ms$) than in the LMF one ($\simeq 45\, \rm ms$); the
breakout also shows signs of intermittency in the case of low magnetic
fields, \eg at $\bar{t} \simeq 110$ or $190\, \rm ms$, mostly because in
this case the stability criterion~\eqref{eq:parker_c} is only mildly
violated. Also different is the strength of the breakout transition,
which is clearly more marked in the case of stronger initial magnetic
fields (right portions in Fig.~\ref{fig:parker_c}). This behaviour has a
clear physical interpretation: given a rather similar dynamics of the
matter, stronger magnetic fields will produce earlier the conditions for
the buoyancy of the low-density matter from the surface of the HMNS. What
discussed so far applies also to the disk region (bottom panels in
Fig.~\ref{fig:parker_c}), although in these scenarios the breakout is
only marginal in the case of the LMF case and is comparatively weaker in
the case of the HMF scenario. A phenomenology of this type can be easily
interpreted when considering that the fluid pressure in the disk is
considerably larger than in the funnel and hence it is harder for the
fluid to breakout from the disk region of the HMNS. We will further
discuss the impact of this breakout in terms of the observable EM
emission in Sec.~\ref{sec:prop_outflow}.

After breakout, the outer layers of the HMNS, \ie with $\rho \lesssim
10^{12}\, \rm{g~cm^{-3}}$, also experience local eruptions of plasma, or
``flares''. These can be appreciated both from the spacetime diagrams in
Fig.~\ref{fig:parker_c} and in Fig.~\ref{fig:flare_view}, which report
vertical sections on the $(x,z)$ plane of the Parker criterion, the inverse
plasma-$\beta$, the fluid specific entropy\footnote{We here use the
specific entropy and not the temperature as it provides a better proxy
of the heating in the low-density plasma.}, and the Lorentz factor;
for each panel, the left and right portions refer to the
LMF and HMF scenarios, respectively.

\begin{figure*}
  \center
  \includegraphics[width=0.98\textwidth]{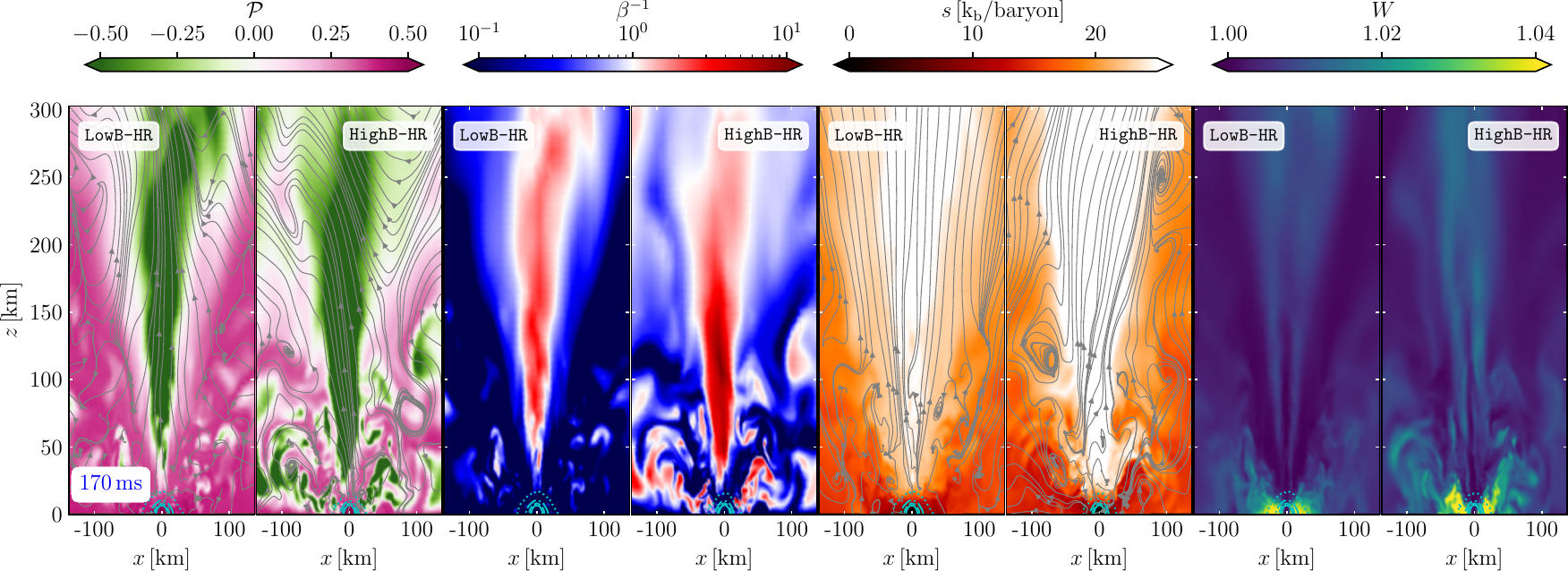}
  \caption{The same as in Fig.~\ref{fig:flare_view}, but showing the
    large-scale structure at time $\bar{t}=170\, \rm ms$. }
	\label{fig:outflow_view}
\end{figure*}
These flares, which have been reported also in other works and
scenarios~\cite{Carrasco2019, Nathanail2020, Nathanail2020c, Most2020b,
  Most2023, Mahlmann23, Musolino2024b}, appear not only in the polar
regions but also near the equatorial plane, enhancing both mass outflow
and angular momentum transport by moving to larger radii matter with
large specific angular momentum. Flares appear in both HMF and LMF cases
but exhibit distinct characteristics: in the HMF case, flares are
stronger and emerge from both low-latitude and polar regions, while in
the LMF case, flares are weaker and predominantly emerge from the polar
region (see Sec.~\ref{sec:prop_flare} for details). These flares help
establish structured magnetic fields and clear the funnel region,
facilitating the development of highly collimated, quasi-steady outflows
(see Sec.~\ref{sec:prop_outflow} for more on their
properties). References~\cite{Zappa2023, Musolino2024b, Ng2024c} suggest
that neutrino fluxes are significantly enhanced along the polar
directions, partly because the equatorial direction is blocked by the
presence of dense matter, and partly because of re-radiation from the
disk after neutrino re-absorption. While neutrinos are not essential for
the breakout and the subsequent flares~\cite{Musolino2024b}, their
inclusion effectively reduces baryon-loading in the polar regions,
leading to more structured, continuous, and strongly collimated
outflows. Although the Poynting fluxes in our simulations are less steady
than those reported in Ref.~\cite{Musolino2024b}, and this is mostly due
to the absence of neutrinos sweeping the polar funnel, strong Poynting
fluxes are still produced in both the LMF and HMF cases (see the
discussion in Sec.~\ref{sec:prop_outflow} below)

Different initial magnetic-field strengths also lead to clear qualitative
differences in the flares, as can be appreciated when comparing the left
(LMF) and right (HMF) portion of each panel in Fig.~\ref{fig:flare_view}.
More specifically, in the HMF scenario, we first observe that flares are
initiated from density regions of $\rho \approx 10^{12}\, \rm g~cm^{-3}$
(see the highly distorted magnetic flux-tube that is a sign of violent
magnetic buoyancy) and further amplified in the density region of $\rho
\approx 10^{11}\, \rm g~cm^{-3}$. Besides, they appear both in the disk
and in the funnel, where matter can be heated to specific entropies of
$\sim 30 \, {\rm k_b}/\text{baryon}$, much comparatively colder matter is
present at lower latitudes. Lastly, we observe some of these low-latitude
flares float upward, enhancing the Poynting flux in the polar region,
while a smaller fraction propagates outward, contributing to low-latitude
emission. By contrast, in the cases with LMFs, flares essentially do not
appear in the disk region and are also much weaker in the funnel (compare
the left and middle panels in Fig.~\ref{fig:flare_view}). Besides, the
weak flares in the funnel region are originated from density regions far
below $\rho \approx 10^{11}\, \rm g~cm^{-3}$. These hold true even when
the magnetic-field strength within the merger remnant of the LMF case
exceeds that of the HMF case, as the strong magnetic field in the LMF
case is mostly confined to regions with densities above $10^{14}\,
\mathrm{g~cm^{-3}}$ before the slow-down stage. It is worth
remarking that the comparison that can be made in
Fig.~\ref{fig:flare_view}, and hence the ability to assess the different
dynamics that characterize the merger remnant under different
magnetic-field strengths, is possible only with a comparative study of
the type presented here. Indeed, had we adopted a single value of the
initial magnetic field, the picture we would have deduced would have
inevitably been a partial one.

By the end of all simulations, and both in the LMF and HMF scenarios, the
merger remnant becomes approximately spherical (see density contours in
Fig.~\ref{fig:flare_view}), partly due to the angular-momentum loss and
partly due to the reduced pinch-effect of the strong toroidal magnetic
field that produces a prolate distribution of
matter~\cite{Frieben2012}. The reduced deformation of the magnetar is
also responsible for a reduction in the amplitude of the remnant
oscillations and for the dynamical ejection of matter, both in the polar
and equatorial directions. As a result, matter will tend to accumulate,
and in the absence of a neutrino wind sweeping away to large distances,
it will lead to a change in the stratification in the outer layers of the
HMNS and hence to a reduced violation of the Parker instability
criterion. In turn, this results into a partial weakening of the Poynting
flux around $\sim 200\, \rm ms$ (see Figs.~\ref{fig:parker_c} and
\ref{fig:outflow}, and the discussion below).

\subsection{Properties of the outflow}
\label{sec:prop_outflow}

\begin{figure*}
  \center
  \includegraphics[width=0.73\textwidth]{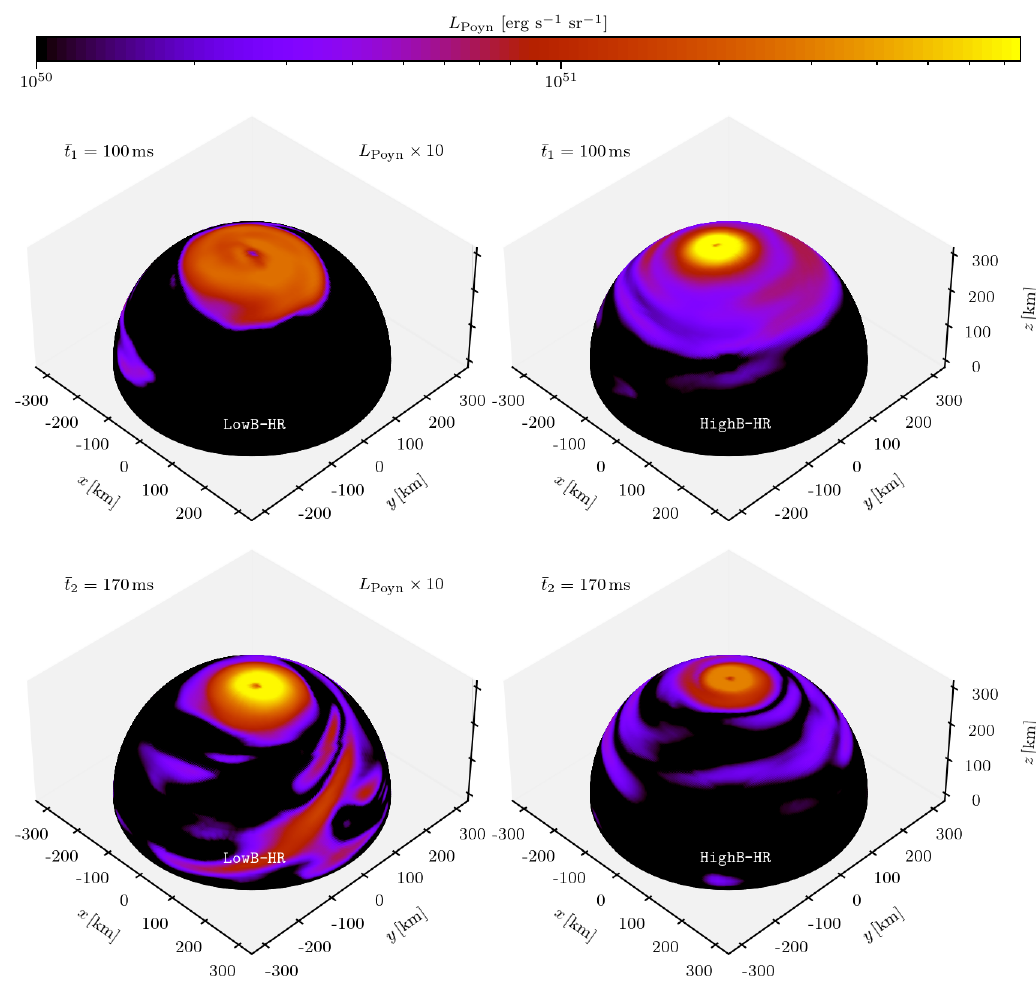}
  \caption{Distributions of the Poynting fluxes as projected on
    two-spheres at a coordinate radius of $300\, \rm km$ at
    $\bar{t}_1=100\, \rm ms$ (top row) and $\bar {t}_2=170\, \rm ms$
    (bottom row), for the low (left panels) and high (right panels)
    initial magnetic fields (LMF and HMF, respectively). Note that in the
    case of LMF scenarios, the Poynting luminosity is multiplied by 10 so
    that it is comparable with the values of the HMF case.}
  \label{fig:outflow_Poynting}
\end{figure*}

Obviously, the Poynting flux represents one of the best tools to obtain
EM information and hence construct a multi-messenger description of the
BNS merger system. As a result, they have been explored, in different
forms and approximations, for almost a decade~(see, \eg
Refs.~\cite{Liu:2008xy, Anderson2008, Rezzolla:2011, Kiuchi2015} for some
of the initial works). We here discuss how to use the Poynting flux to
distinguish between the long-term evolutions with low and high initial
magnetic fields. We start by reporting in Fig.~\ref{fig:outflow_view} the
two-dimensional properties of the outflow on the $(x,z)$-plane in terms
of the same quantities used in Fig.~\ref{fig:flare_view}, namely, from
left to right, the Parker criterion, the inverse plasma-$\beta$, the
specific entropy, and the Lorentz factor. To emphasise the large-scale
structure of the collimated outflows, we select $\bar {t} = 170\,
\text{ms}$, the point at which the Poynting flux in the LMF case reaches
its peak across the entire simulation.

Figure~\ref{fig:outflow_view} clearly shows that highly collimated outflows
form in the polar regions of both the LMF and HMF cases and are
characterized by ordered and large-scale poloidal magnetic-field lines
that serve as guide for the motion of the fluid. These outflows are also
characterized by high values of the inverse plasma-$\beta$, significant
heating, but only moderate Lorentz factors. More specifically, in the HMF
(LMF) case, the inverse plasma-$\beta$ can readily exceed values
$\mathcal{O}(10)$ ($\mathcal{O}(5)$) within the funnel region, leading to
a strong Parker instability, which transports highly magnetized material
upwards. Furthermore, the Lorentz factors measured at about $300\,{\rm
  km}$ are only $W = 1.02$ for the HMF scenario, to which correspond
asymptotic values $W_{\infty} := - h u_t = 1.17$ that are much smaller
than those expected in the phenomenology of short GRBs. Finally, the
density contours reveal that the merger remnant expands in the vertical
($z$-axis) direction, with matter accumulating in this region. This
accumulation contributes to the eventual weakening of the strong Poynting
flux (see also Sec.~\ref{sec:parker}).

In order to gain a full 3D view of the propagation of these outflows, we
have analyzed their properties via the projection of the Poynting flux on the
two-sphere with coordinate radius $r\simeq 300\,{\rm km}$. In particular,
we show in the top panel of Fig.~\ref{fig:outflow_Poynting} the
projection at $\bar{t}=100\, \rm ms$, while the bottom panel refers to a
much later time of $\bar {t}=170\, \rm ms$. Furthermore, for each panel,
the left and right parts refer to the LMF and HMF scenario,
respectively; note that because the corresponding EM luminosities are
considerably different in the LMF and HMF, we have multiplied by a factor
of ten the Poynting luminosity of the LMF to make it appear on the same
colormap.

\begin{figure}
  \center
  \includegraphics[width=0.45\textwidth]{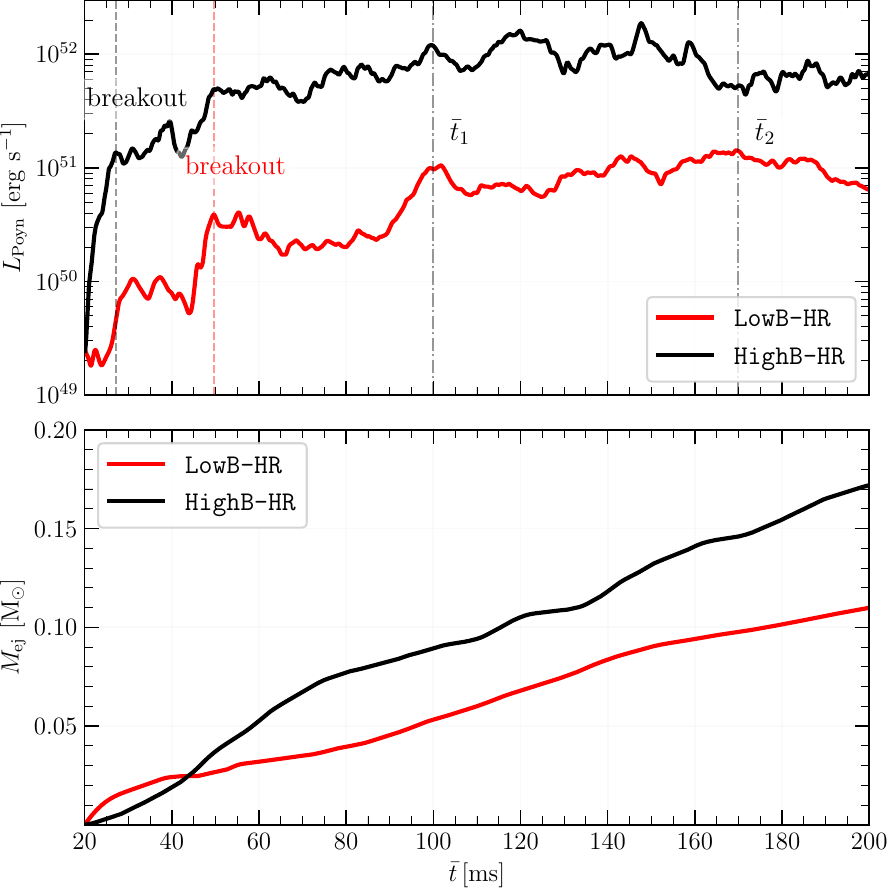}
  \caption{Evolution of the Poynting luminosity (top panel) and of the
    cumulative unbound mass (bottom panel) as computed on a two-sphere at
    a radius of $300\, \mathrm{km}$ for the LMF (red solid line) and the
    HMF (black solid line). Shown with dashed lines are the breakout
    times in the two scenarios, while the dot-dashed lines mark the 
    times corresponding to the outflow properties shown in
    Fig.~\ref{fig:outflow_Poynting} and Fig.~\ref{fig:outflow_Poynting_1}.}
  \label{fig:outflow}
\end{figure}

\begin{figure}[t!]
  \center
  \includegraphics[width=0.42\textwidth]{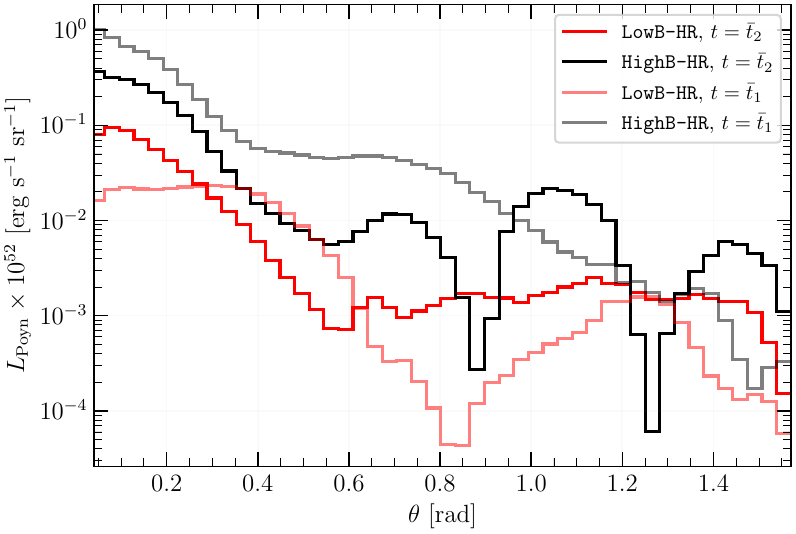}
  \hskip  0.25cm
  \includegraphics[width=0.42\textwidth]{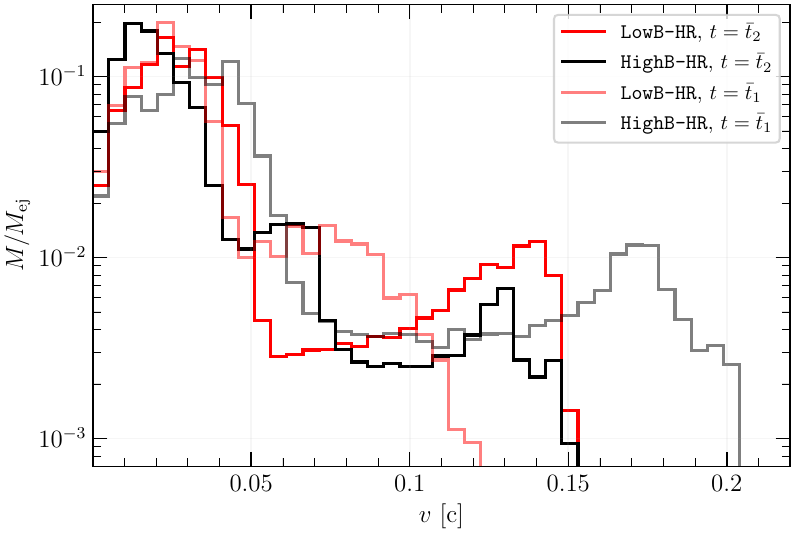}
  \caption{Angular distribution of the Poynting flux (top panel) and
    velocity distribution of the unbound matter (bottom panel) as
    measured by a detector at radius of $300\, \rm km$. Lines of
    different colours refer to either the LMF (red line) or to the HMF
    (black line) scenario. Furthermore, lines of different shadings show
    the distributions at different times, namely at $\bar{t}_1=100\, \rm
    ms$ (lighter lines) or at $\bar{t}_2=170\, \rm ms$ (darker lines), so as
    to show how these quantities evolve in time (see also
    Fig.~\ref{fig:outflow_Poynting} and Fig.~\ref{fig:outflow}).}
  \label{fig:outflow_Poynting_1}
\end{figure}

Overall, Fig.~\ref{fig:outflow_Poynting} allows us to appreciate that in
the LMF case, the Poynting flux is not highly collimated at
$\bar{t}=100\, \rm ms$, though it remains confined within a region with a
polar angle of $\theta < \pi/3$. As the outflow evolves, it becomes
increasingly collimated, with the Poynting flux being eventually confined
within a narrower region of $\theta < \pi/6$ at the time of its peak
value, which occurs at $\bar{t}=170\, \rm ms$ (see also
Fig.~\ref{fig:outflow}). In the HMF scenario, on the other hand, the
Poynting flux is consistently stronger, resulting from earlier breakup
and a more extended distribution of magnetic-field structures, along with
disk contributions (see also Sec.~\ref{sec:parker}) and strong flares
from higher density regions (see also Sec.~\ref{sec:prop_flare}).
Interestingly, at $\bar{t} = 170 \, \rm{ms}$, the angular distributions
of the Poynting flux in the polar region are remarkably similar in both
the LMF and HMF cases. The primary distinction lies in the magnitude,
which is larger by a factor of $\mathcal{O}(4)$ when strong initial
magnetic fields are present. In contrast, at lower latitudes, multiple
peaks associated with low-latitude flares are observed in the HMF
case (see also Fig.~\ref{fig:outflow_Poynting_1}). However, no such
strong peaks appear in the LMF case, where
the angular distribution of the Poynting flux remains comparatively
flat. Interestingly, the rotation of the merger remnant and the flares
that take place in its outer layers, are clearly imprinted in the
fluctuations that are visible on the two-sphere projections. Hence, this
phenomenology opens the possibility of learning about the rotation rate
of the remnant by carefully studying the time statistics of these
fluctuations in those (possibly rare) cases where they are measurable.

We report a more quantitative measurement of the EM emission from the
remnant in the upper panel of Fig.~\ref{fig:outflow}, which shows the
Poynting luminosity as computed starting from the time when the Poynting
flux reaches a two-surface at $300\, \mathrm{km}$ using the
\texttt{HEALPix} discretization~\cite{Gorski2005} (see
Appendix~\ref{sec:post_eqs} for details). Note that the Poynting fluxes in
the HMF case are from $4~\text{to}~10$ times stronger than those in the
LMF case (upper panel of Fig.~\ref{fig:outflow}) and that the peak
luminosity in the HMF case reaches $1.9 \times 10^{52}\, \rm erg~s^{-1}$,
while it is $1.4 \times 10^{51}\, \rm erg~s^{-1}$ in the LMF
scenario. Additionally, the HMF (LMF) case sustains a continuous Poynting
flux of over $5 \times 10^{51}\, \rm erg~s^{-1}$ ($5 \times 10^{50}\, \rm
erg~s^{-1}$) for $100\,\rm ms$ ($110\,\rm ms$).

\begin{figure*}
  \center
  \includegraphics[width=0.9\textwidth]{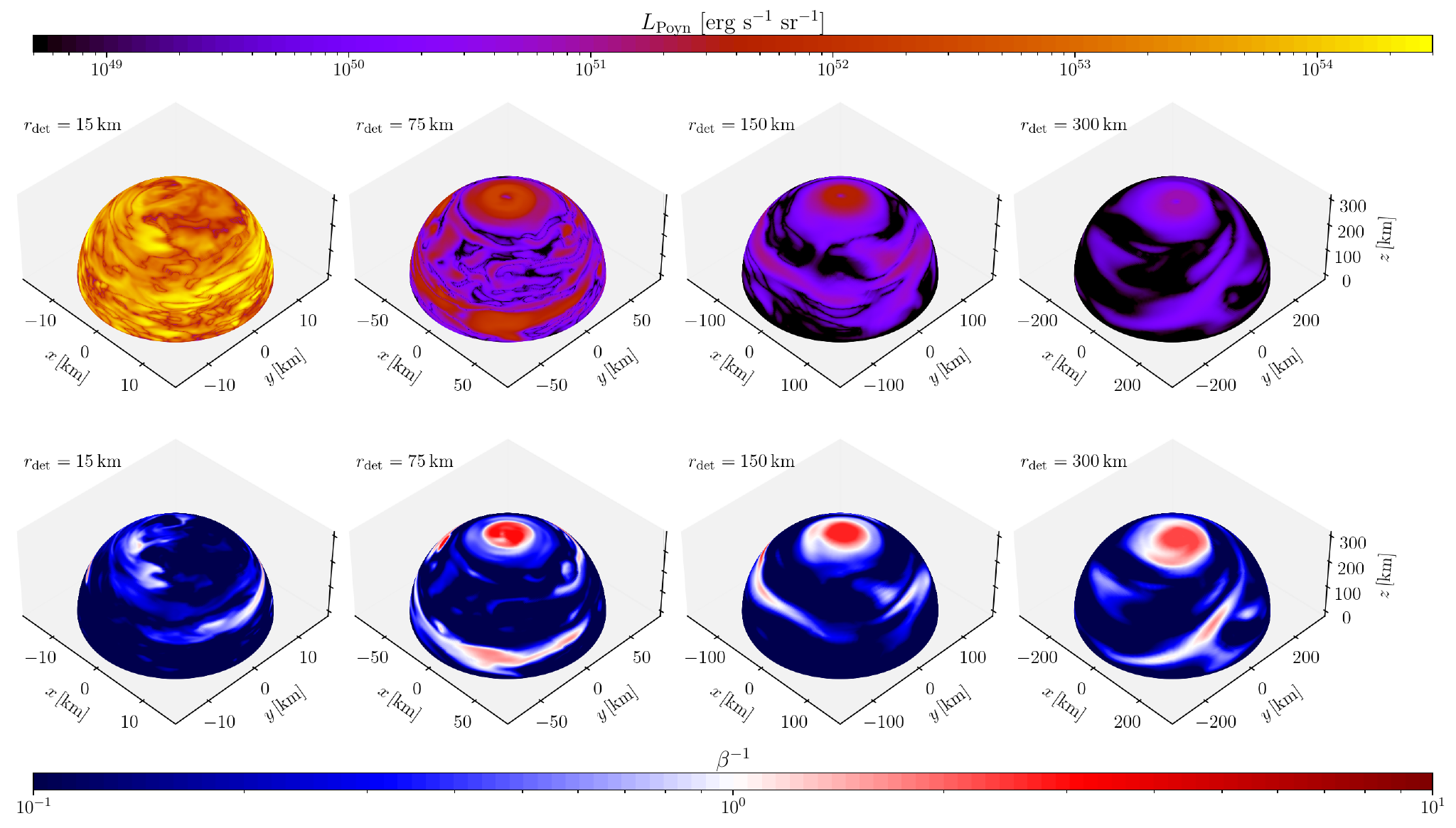}
  \caption{Distributions of the Poynting fluxes (top row) and of the
    inverse plasma-$\beta$ (bottom row) as projected on two-sphere of
    four different radii (\ie from left to right: $15\, \rm km$, $75\,
    \rm km$, $150\, \rm km$, and $300\, \rm km$) at $\bar{t}_2=170\, \rm
    ms$. The data refers to the LMF case; a similar evolution is shown in
    Fig.~\ref{fig:dynamic_flare_highB} for the HMF case.}
  \label{fig:dynamic_flare_lowB}
\end{figure*}

All things considered, the differences between the EM luminosity in the
HMF and LMF cases can be ascribed to three main factors. First, the
earlier breakout of the magnetic field in the HMF case allows for the
propagation of a larger amount of EM energy from the remnant, cleaning up
the funnel; by contrast, in the LMF case the EM energy remains confined
close to the remnant for a longer time. Second, in the HMF case, flares
originate from a higher-density region of the magnetar, causing stronger
eruptions than those produced in the LMF case, where flares are rooted in
the relatively low-density region. Third, the Parker instability
criterion is frequently met in the low-latitude regions of the HMF
scenario (see bottom panel of Fig.~\ref{fig:parker_c}), contributing to
stronger Poynting flux; on the contrary, the criterion is marginally met
and only in the funnel in the LMF case. Interestingly, a stronger
Poynting flux produced by a stronger magnetic field was reported also by
Ref.~\cite{Bamber2024b}, though their simulations were conducted on a
much shorter timescale.

A more quantitative measure of the anisotropy of the emission and of the
degree of collimation of the outflows is presented in top panel of
Fig.~\ref{fig:outflow_Poynting_1}, which reports the polar distribution
of the Poynting luminosity in the case of the LMF (red solid lines) and
of the HMF (black solid lines) scenarios; furthermore, different shadings
refer either to an early time at $\bar{t}_1=100\, \rm ms$ (lighter lines)
or to a late one at $\bar {t}_2=170\, \rm ms$ (darker lines). Concentrating
on the latter, it is clear that the degree of collimation is the same in
the LMF and HMF cases, but that the intensity of the luminosity differs
by an order of magnitude. Also quite evident when comparing the
distributions at different times is that the degree of collimation
increases with time, as the merger remnant reaches a quasi-stationary
evolution and the fluctuations related to the ejection of matter have
been washed out.

Also reported in Fig.~\ref{fig:outflow}, but in the lower panel, is the
evolution of the ejected mass (see Appendix~\ref{sec:post_eqs} for
details on the definition and calculation). Clearly, a substantial amount
of mass is ejected by the remnant over the timescale of the simulation
and this amounts to a total of $M_{\rm ej} \sim 0.11\, M_{\odot}$ for the
LMF case and $\sim 0.17\, M_{\odot}$ for the HMF case, thus indicating
that strong magnetic fields in the merger remnant greatly facilitate the
mass ejection. Less obvious is why for $20\lesssim \bar{t} \lesssim
50\,{\rm ms}$ the ejected mass is larger in the LMF case than in the HMF
case. We believe that this is because of the larger braking action in the
HMF case, that converts the kinetic-energy reservoir in the remnant into
toroidal magnetic field. This braking is smaller for the LMF scenario and
hence the kinetic energy is employed more efficiently to eject unbound
matter. Part of the reason may stem from the pinch effect of the strong
toroidal magnetic field (see, \eg~\cite{Frieben2012}), which has already
reached very large values by $\bar{t} \sim 20\, \rm{ms}$ in the HMF
case. This effect can limit matter expansion in the disk region, the
primary source of the ejected mass.

Finally, shown in the bottom panel of Fig.~\ref{fig:outflow_Poynting_1}
is the velocity distribution of the unbound mass with the same colour
convention used for the distribution of the Poynting luminosity in the
top panel. Note that for both magnetic-field configurations, the
distribution reveals two components: a subdominant high-velocity
component, \ie with $0.1 \lesssim v \lesssim 0.15$, associated with the
matter ejected from the polar region, and a dominant low-velocity
component, \ie with $v \lesssim 0.1$, primarily contributed by material
from the disk.

\begin{figure*}
  \center
  \includegraphics[width=0.88\textwidth]{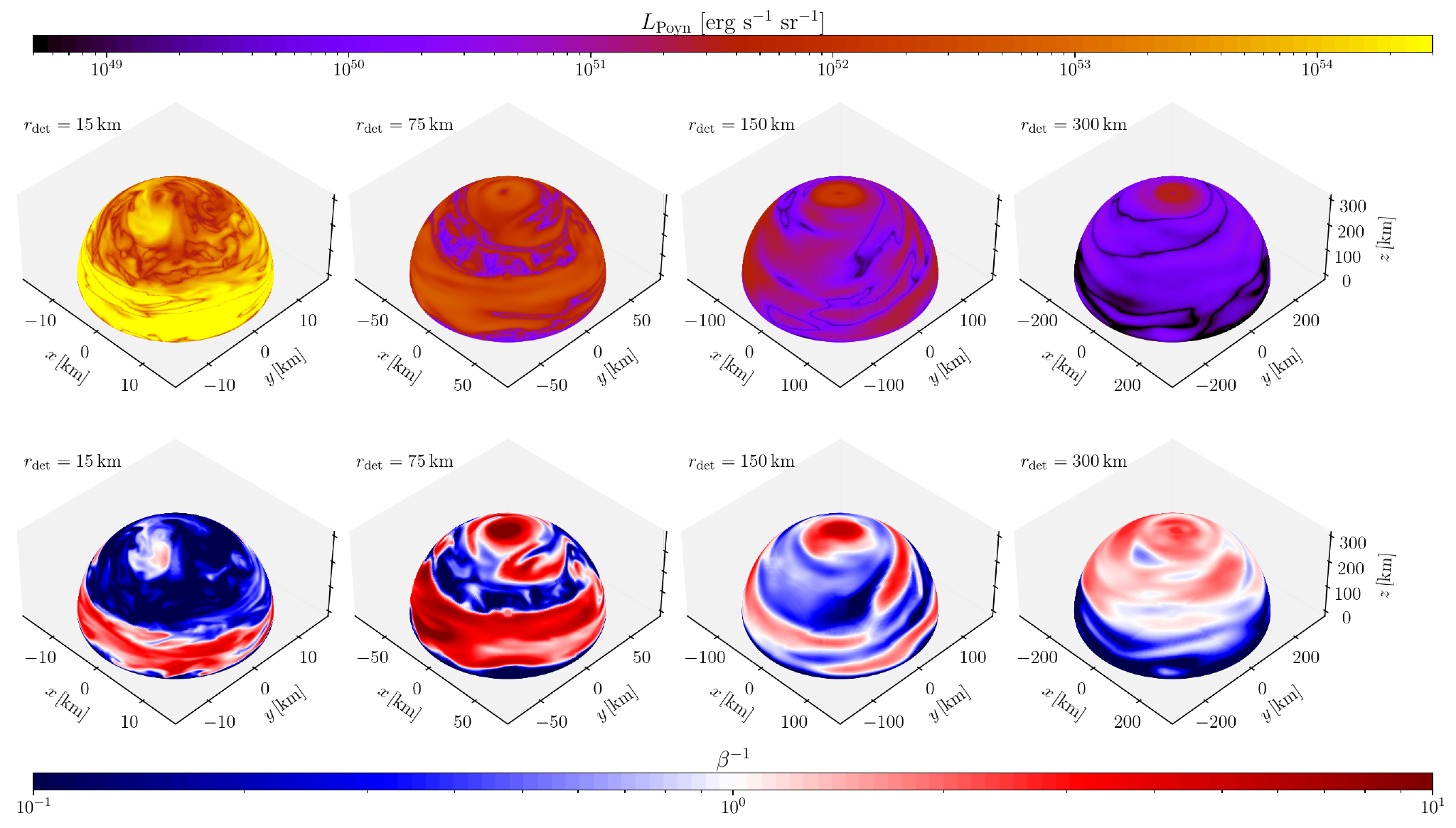}
  \caption{The same as in Fig.~\ref{fig:dynamic_flare_lowB}, but for the
    HMF case.}
  \label{fig:dynamic_flare_highB}
\end{figure*}

Note also that the late-time distribution of the HMF case has lost part
of its high-velocity tail exhibited at $\bar{t}_2 = 170\, \mathrm {ms}$
and that reached values $v \lesssim 0.2$. This is because the most
energetic emission in the case of large initial magnetic fields takes
place early on and for $\bar{t} \lesssim 120\, \mathrm
{ms}$. Interestingly, the opposite takes place for the LMF case, where
the late-time distribution gains a high-velocity component. This is
because the LMF Poynting flux reaches its peak value at $\bar{t}_2 =
170\, \mathrm {ms}$, as buoyant material driven by Parker instability
clears the funnel region, allowing ejected matter to escape more rapidly.
However, the corresponding Lorentz factors in all cases remain rather low
and smaller that those reported in Refs.~\cite{Kiuchi2023, Most2023,
  Musolino2024b}, as the funnel region is heavily polluted by baryons,
and neutrino emission is neglected in the simulation. Including the
latter will boost the material to larger relativistic velocities with
terminal Lorentz factors as large as $2-20$~\cite{Kiuchi2023,
  Musolino2024b}, which however fall short of the values
$\mathcal{O}(100)$ expected from the phenomenology of short
GRBs~\cite{Duncan1992, Usov1992, Dai1998, Zhang2001}.

\subsection{Dynamics of the flares near the HMNS}
\label{sec:prop_flare}

In the previous Sec.~\ref{sec:parker}, we demonstrated that flares are
generated by the Parker instability and analyzed their formation sites.
Here, we extend our analysis to their dynamics in the vicinity of the
HMNS. As shown in Fig.~\ref{fig:flare_view}, the Parker instability gives
rise to magnetic-field arc structures, which are further inflated by
convective motion. Due to the highly differential rotation of the
remnant, the footpoints of these arcs undergo differential motion,
leading to the formation of twisted magnetic loops. These loops rise as a
result of magnetic buoyancy, transporting magnetically dominated material
upward. This process heats the surrounding matter and drives a strong
Poynting flux, introducing additional variability in the EM
emission. Furthermore, these eruptions play a crucial role in shaping the
structured poloidal magnetic field in the polar region (see
Sec.~\ref{sec:prop_outflow}). This configuration facilitates mass outflow
and angular momentum transport along the $z$-direction via the
magnetocentrifugal mechanism.

To further quantify these dynamical processes, we present in
Figs.~\ref{fig:dynamic_flare_lowB} and~\ref{fig:dynamic_flare_highB} a
series of projections on two-spheres in the LMF and HMF cases,
respectively. More specifically, for each figure, in the top part we show
the projections of the Poynting luminosity on different extraction radii,
while in the bottom one we report the corresponding projections for the
inverse plasma-$\beta$. We should note that capturing the evolution of
these flares and of their highly dynamical propagation is inherently
challenging; thus, our approach provides an indirect yet insightful
perspective on the underlying dynamics. Notably, we find that eruptions
near the HMNS significantly enhance the Poynting flux in both the LMF and
HMF cases. This enhancement is evident as regions with large $\beta^{-1}$
consistently exhibit a higher Poynting flux than their surroundings.

At small radii, \ie $r \sim 15\, \rm{km}$, the strong Poynting flux
primarily originates from flares, namely, matter outflows carrying
intense magnetic fields during the violent oscillation phase of the HMNS.
At the eruption, magnetic energy in the flare is converted to kinetic
energy, but as they propagate outward, they naturally interact with the
ambient plasma, transferring part of their kinetic energy, but also
expanding and cooling, and ultimately weakening their strength, as a
result, the Poynting flux at large radii, \ie $r \sim 300\, \rm{km}$, has
been reduced significantly. Furthermore, we observe distinct
morphological differences across the different two-spheres: in the polar
regions, structures tend to form ring-like patterns, whereas at lower
latitudes, they resemble belts. This distinction arises naturally due to
the rapid rotation of the remnant and the large degree of anisotropy in
the ejection of matter. 

Obviously, together with this common qualitative phenomenology, the
flares in the LMF and HMF scenarios measured on the two-spheres differ
quantitatively in the frequency and strength becoming both more and
stronger as the initial magnetic field is increased, as already
encountered with Figs.~\ref{fig:parker_c} and \ref{fig:flare_view}. This
is not surprising and indeed it is consistent with the expectation that
flares simply reflect the breaking of a local hydromagnetic equilibrium
in the outer layers of the merger remnant. As such, stronger magnetic
fields naturally favour the breaking of the local hydromagnetic
equilibrium.

In summary, these intermittent eruptions from the HMNS, that have been
reported in other works and scenarios~\cite{Siegel2014, Kiuchi2017,
  Most2020b, Most2023, Most2023b, Carrasco2019, Nathanail2020,
  Nathanail2020c, Mahlmann23, Musolino2024b}, contribute to strengthen
the Poynting flux, to the formation of a structured poloidal magnetic
field in the polar region, and to aiding collimated outflow
formation. These considerations hold true even if the magnetic fields are
not very strong, but the frequency and strength of the flares is
obviously closely related to the intensity of the magnetic energy. If
the flares reported in this work were to be sustained over long periods
of time and detected in the emission before and after a short (or even a
long) GRB, they could reveal the presence of a highly magnetized,
long-lived neutron star following a BNS merger (see also relevant
discussion in Refs.~\cite{Dai1998b, Fan2006, Gao2006, Dai2006,
DallOsso2011, Rowlinson2013, Wang2024}). In turn, this information
could be used to further constrain the maximum neutron star mass and EOS
and some intriguing observational evidence has possibly already
emerged~\cite{Chirenti2023, Yang2025}. We note that, aside from the MHD
effects identified by Refs.~\cite{Siegel2013, Combi2023, Kiuchi2023,
  Most2023b, Musolino2024b}, which suggest that the MRI and dynamo
processes can drive relativistic outflows or flaring, our results reveal
that the Parker instability can directly or indirectly contribute to
flaring and relativistic outflows both before and after a short
GRB. Clearly, future simulations containing full neutrino and EM
radiative transfer process are needed to sharpen this picture and obtain
more precise predictions on the role played by the Parker instability and
by magnetic flares on the phenomenology of BNS mergers, but also of long
GRBs.

\section{Conclusions}
\label{sec:summary}

Much of the future progress in the modelling of the merger of binary
systems comprising at least a neutron star will depend very strongly on
the ability to carry out simulations that model the system, either in the
inspiral or after the merger, for timescales that are
$\mathcal{O}(100-1000)$ times the dynamical one. Such timescales,
however, are prohibitively expensive for the standard approaches in
numerical relativity to simulate BNS and BHNS mergers.

Our approach to tackle these challenges consists in a hybrid approach
that we have recently developed and that combines, via a hand-off
transition, a fully general-relativistic code, \ie \FIL, with a more
efficient code, \ie \BHAC, making use of the conformally flat
approximation (CFC). A crucial aspect of the CFC approach that makes it
computationally attractive is that the corresponding field equations are
elliptic and need not to be updated at every time-step over which the
GRMHD equations are solved. Rather, they can be solved every $10$-$40$
time-steps, thus decreasing computational costs. In addition, when
solving the set of GRMHD equations, the time-step size needs not be
CFL-limited by the speed of light, but by the relevant MHD speeds, which
are $\sim 30\%$ smaller. As a result, speed-ups of the order of $3-4$ are
possible in our hybrid approach and, in turn, simulations on timescales
$\mathcal{O}(100-1000)\,{\rm ms}$ can be carried out with much reduced
computational costs.

While our hybrid approach was first presented and discussed in detail in
Ref.~\cite{Ng2024b}, we have here reported important additional
developments of \BHAC consisting of the inclusion of gravitational-wave
radiation-reaction contributions and of higher-order formulations of the
equations of GRMHD. Both improvements have allowed us to explore the BNS
merger remnants with high accuracy and over timescales that would have
been computationally prohibitive with \FIL.

More specifically, using these improvements, we have stress-tested the
hybrid approach carrying out the \HO procedure under different conditions
of the spacetime dynamics and GW content. Our findings indicate that this
approach yields highly accurate evolutions for the postmerger remnant,
capturing both overall dynamics and magnetic field evolution
effectively. The agreement between the two codes greatly improves as the
GW amplitude decreases, especially when GW amplitude has decreased to
$1/16$ of its peak value. We suggest the latter as the reference
criterion for performing the \HO as it is gauge-invariant, model
independent, and only weakly dependent on resolution.

Besides reporting computational developments, we have applied the new
hybrid approach to investigate the impact of the magnetic-field strength
on the long-term and high-resolution evolutions of the ``magnetar''
resulting from the merger of two neutron stars with a realistic EOS.
More specifically, using the TNTYST EOS, we have simulated magnetars
formed after BNS mergers exploring two different magnetic-field strengths
representing a low and a high-magnetic-field scenario (LMF and HMF,
respectively) with a resolution of $150\, \rm m$ over a simulation period
of $200\, \rm ms$. Interestingly, in the LMF case, the winding time is
prolonged, and the magnetic field generated through winding is eventually
larger than that of the HMF case. This behaviour can be attributed to a
stronger back-reaction of the toroidal magnetic-field tension and hence
to the suppression of small-scale turbulence in the high magnetic-field
scenario. Furthermore, while in both scenarios we observe the weakening
or suppression of differential rotation, the latter takes place on
smaller timescales when the magnetic field is stronger, as expected.

Interestingly, we find that the physical conditions in the outer layers
of the merger remnant close to the polar axis can lead to a global
breakout of the plasma as a result of the development of the Parker
instability (as also reported in Ref.~\cite{Musolino2024b}). This
breakout leads not only to an enhancement of the Poynting flux, but also
to the occurrence of local eruptions of plasma, or ``flares'', that
are present in both the LMF and HMF scenario, although with different
frequency and strengths. These flares, that are driven mostly by the
Parker instability and have been reported in other works and
scenarios~\cite{Siegel2014, Kiuchi2017, Most2020b, Most2023, Most2023b,
  Carrasco2019, Nathanail2020, Nathanail2020c, Mahlmann23,
  Musolino2024b}, contribute to strengthen the Poynting flux, to the
formation of a structured poloidal magnetic field in the polar region,
and to aiding collimated outflow formation. As a result, we measure
Poynting luminosities reaching $1.9 \times 10^{52}\, \rm erg~s^{-1}$
($1.4 \times 10^{51}\, \rm erg~s^{-1}$) and collimated within a
half-opening angle of $\pi/6$ for the HMF (LMF) scenario. Independent
of the strength of the initial magnetic field, the growth of the magnetic
energy slows down and the main difference is in the time this takes
place, which can be $\mathcal{O}(30)\,{\rm ms}$ for the HMF or
$\mathcal{O}(100)\,{\rm ms}$ for the LMF. Hence, by the time the
simulations are ended after $\sim 200\,{\rm ms}$ the collimated Poynting
fluxes are weakened. The inclusion of neutrino transport, and the
consequent clearing of the funnel operated by the neutrino-driven winds,
are likely to counteract the decrease in Poynting flux and possibly
increase the impact of the flares on the variability of the EM
luminosity. 

As a final remark we note that if these eruptions were to be sustained
over long timescales and detected in the emission before and after a
short GRB, they could reveal the presence of a long-lived magnetar in a BNS
merger and potentially provide information of the properties of the
remnant and its EOS. While these prospects are very exciting and some
intriguing observational evidence may have already
emerged~\cite{Chirenti2023, Yang2025}, future simulations are needed to
confirm this picture. In particular, including a treatment of neutrino
and EM radiation transfer and carrying out these simulations on even
longer timescales, will certainly help attain a more realistic
description of the long-term evolution of the remnant from BNS
mergers. We plan to report on these improvements in future work.

\begin{acknowledgments}
We thank R. Oechslin and G. Schaefer for the extensive exchange on the
inclusion of the RR terms and E. Most and C. Musolino for useful
discussions. This research is supported by the ERC Advanced Grant
``JETSET: Launching, propagation and emission of relativistic jets from
binary mergers and across mass scales'' (grant No. 884631), by the
Deutsche Forschungsgemeinschaft (DFG, German Research Foundation) through
the CRC-TR 211 ``Strong-interaction matter under extreme conditions'' --
project number 315477589 -- TRR 211, by the GSI Helmholtzzentrum f\"ur
Schwerionenforschung, Darmstadt as part of the strategic R\&D
collaboration with Goethe University Frankfurt and by the State of Hesse
within the Research Cluster ELEMENTS (Project ID 500/10.006). JLJ
acknowledges partial support by the Alexander von Humboldt Foundation.
MC acknowledges support from the NSF grants PHY-2110338, PHY-2409706,
AST-2031744 and OAC-2004044 as well as the NASA TCAN Grant
No. 80NSSC24K0100. LR acknowledges the Walter Greiner Gesellschaft zur
F\"orderung der physikalischen Grundlagenforschung e.V. through the Carl
W. Fueck Laureatus Chair. The simulations were performed on HPE Apollo
HAWK at the High Performance Computing Center Stuttgart (HLRS) under the
grant BNSMIC.

\paragraph{Software.~} The software employed in this work comprises
the following codes: \texttt{BHAC}~\cite{Porth2017, Olivares2019,
  Ripperda2019}, \texttt{BHAC+}~\cite{Ng2024b},
\texttt{FIL}~\cite{Most2019b, Chabanov2022, Musolino2023, Chabanov2023,
  Ng2024c}, \texttt{ETK}~\cite{Loffler:2011ay},
\texttt{FUKA}~\cite{Papenfort2021b}. Furthermore, it makes use of the
\texttt{CompOSE}
(\href{https://compose.obspm.fr}{https://compose.obspm.fr}) database for
the handling of the EOS~\cite{Typel2015}.

\end{acknowledgments}

\appendix

\section{Diagnostic quantities}
\label{sec:post_eqs}

In what follows we provide some explicit expressions for the definition
of quantities discussed extensively in the main text. Different
implementations of the same formula may be used in \FIL and \BHAC because
different gauge conditions and metric forms are adopted in these two
codes.

We start with the baryon mass that is calculated as
\begin{equation}
    \label{eq:bary_mass}
     M_{\rm b} :=  \int_{\mathcal{V}} W \rho \sqrt{\gamma}\, d^3\,x\,,
\end{equation}
where $W$ is the Lorentz factor of the fluid, and $\gamma$ is the
determinant of the purely spacial metric. Similarly, the total internal
energy is defined by
\begin{equation}
  \label{eq:int}
  E_{\rm int} :=  \int_{\mathcal{V}} W\epsilon \rho \sqrt{\gamma}\, d^3\,x\,,
\end{equation}
where $\epsilon$ is specific internal energy. Another important energy
considered in our analysis the total EM energy that in \BHAC is defined
as
\begin{equation}
  \label{eq:em_total}
  E_{\rm EM} := \frac{1}{2} \int_{\mathcal{V}} (B_i B^i+E_i E^i)
  \sqrt{\gamma}\, d^3\,x\,,
\end{equation}
where $B^i$ and $E^i$ are the magnetic and electric fields measured in
the Eulerian frame, respectively. We note that the equivalent expression
in \FIL contains a multiplicative factor $1/4\pi$ which comes from the
different EM units used in the two codes~\cite{Most2019b, Porth2017}. In
addition, the EM energy in the toroidal magnetic field is computed as
\begin{equation}
  \label{eq:em_tor}
  E_{\rm EM, tor} := \frac{1}{2} \int_{\mathcal{V}} (B_\phi B^\phi+E_\phi
  E^\phi) \sqrt{\gamma}\, d^3\,x\,,
\end{equation}
while the poloidal EM energy is calculated as the difference between the
total and toroidal EM energies. Finally, since we consider the ideal-MHD limit,
the following relation holds
\begin{equation}
  \label{eq:electricf}
  E^i= \gamma^{-1/2} \eta^{ikl}  v_k B_l\,,
\end{equation}
where $v_k$ is the Eulerian velocity, and $\eta^ {ikl}$ is the
three-dimensional Levi-Civita symbol defined as
\begin{align}
\eta^{ikl}=\eta_{ikl} :=\left\{\begin{array}{ccc}
+1 & \mathrm{if} & (i,k,l) \mathrm{~even~permutation} \,,\\
-1 & \mathrm{if} & (i,k,l) \mathrm{~odd~permutation} \,, \\
0  & \mathrm{if} & \mathrm{else}\,.
\end{array}\right.
\end{align}

In terms of radiative quantities, the Poynting flux is calculated on a
two-sphere of coordinate radius $r$ as
\begin{equation}
    \label{eq:poynting}
    L_{\rm Poyn} := \oint_{dS} \left( b^{\mu}b_{\mu} u^r u_0- b^r
    b_0\right) \alpha \psi^6 r^2 \sin{\theta} d\phi d\theta\,,
\end{equation}
where $b^{\mu}$ is the magnetic field in the fluid frame and $u^{\alpha}$
is the four-velocity in the Eulerian frame. The unbound rest-mass is
calculated in a similar way using
\begin{equation}
  \label{eq:ubmass}
  \dot{M}_{\rm ej} = -\oint_{dS} W\tilde{\rho} \left(\alpha v^r-\beta^r \right) \psi^6
  r^2 \sin{\theta} d\phi d\theta\,,
\end{equation}
where $\tilde{\rho}$ refers to matter fulfilling the Bernoulli criterion
$hu_t<-h_{\rm min}$, where $h_{\rm min}$ is the minimum value of specific
enthalpy available in the tabulated EOS~(see, \eg~\cite{Bovard2016} for
a discussion). The surface integral in Eq.~\eqref{eq:ubmass} is performed
using the \texttt {HEALPix} discretization~\cite{Gorski2005} of a
two-sphere to facilitate the calculation of the flux of matter, which has
revealed to be particularly effective in computing both outward-bound and
inward bound fluxes, as those related to fall-back
accretion~\cite{Musolino2024}. In our implementation we employ a
discretization parameter $N_{\rm side}=128$, which corresponds to a size
of $27^{\prime}\,30^{\prime\prime}\times27^{\prime}\,30^{\prime\prime}$
for each pixel on the sphere~\cite{Gorski2005}. The convergence of the
fluxes with respect to $N_{\rm side}$ has been checked and the default
$N_{\rm side}$ is chosen as a balance between angular resolution and the
computational cost.

\section{Details on the inclusion of RR terms}
\label{sec:GWBRderivation}

Although the first inclusion of RR terms was presented in
Ref.~\cite{Oechslin07a}, a clear derivation of the~PN corrections and
their exact coupling to a GRMHD code with CFC approximation has not
been presented yet. In addition, since unreported typos~\cite{Oechslin2024}
appear in Eq.~(A.37) of Ref.~\cite{Oechslin07a}, we
report briefly the expressions for the PN-corrected components of the
metric that are implemented in \BHAC and refer the reader to
Refs.~\cite{Blanchet90, Faye2003} for the full derivations, with the
caveat that these references derive quantities to be implemented within
purely PN numerical codes rather than in codes that solve the
equations of GRMHD or that adopt the CFC approximation.

The most important RR correction in our CFC approach is contained in
Eq.~(\ref{eq:g00_short}) for the metric component $g_{00,{_{\rm RR}}}$
and to discuss how this is obtained we first need to recall the 2.5~PN
order formalism with 3.5~PN corrections derived in Ref.~\cite{Faye2003}
and which improves upon the formalism in Ref.~\cite{Blanchet90}. For
convenience, we will report explicitly the powers of $c$ in the PN
expansion and follow the notation of Ref.~\cite{Faye2003} if the
variables are not defined in this paper, where the metric $g_{\mu\nu}$
splits into an odd $(g_{\mu\nu})_{({\rm odd})}$ and an even
$(g_{\mu\nu})_{({\rm even})}$ part depending the parity of the terms they
generate in the equations of motion of PN hydrodynamics. For example,
$(g_{00})_{(7)}$, $(g_{0i})_{(6)}$ or $(g_{ij})_{(5)}$ belong to the odd
part, while $(g_{00})_{(2)}$, $(g_{0i})_{(3)}$ or $(g_{ij})_{(2)}$ to the
even one; note that both the odd and even parts include 2.5 and 3.5~PN
corrections (see~\cite{Faye2003} for more details).

We start therefore with considering the components of the RR part of the
gravitational field at 2.5~PN and 3.5~PN orders: $(g_{00})_{ (7), {_{\rm
      RR}}}$ and $(g_{00})_{(9), {_{\rm RR}}}$ [see Eqs.~(4.10a) and
  (4.18a) in~\cite{Faye2003}]. Given the definitions of the PN 
potentials $\mathcal{U}_*$, $\mathcal{U}_{* i}$, and $\mathcal{R}$ in
Eqs.~(\ref{eq:PNcorr1})--(\ref{eq:PNcorr3}), and considering only
$Q^{[3]}_{ij}$ related terms, we follow~\cite{Oechslin07a} and replace
$D^*$ with the 1~PN + 3.5~PN mass density, \ie
\begin{equation}\label{eq:sigma}
\sigma := T^{00} + T^{ii}\,,
\end{equation}
and replace $v^i$ with the covariant momentum per unit rest-mass in the fluid
frame
\begin{equation}
w_i := hu_i = v^i + \mathcal{O}(1/c^2)\,, 
\end{equation}
where it includes the 1~PN correction of $v^i$.

Applying these replacements and keeping leading order terms with dependency
of $Q^{[3]}_{ij}$,
the expression for the $(g_{00})_{(7), {_{\rm RR}}}$
metric component can be simplified as follows
\begin{equation}\label{eq:g7}
\begin{aligned}
\left(g_{00}\right)_{(7), {_{\rm RR}}}&=\frac{4 }{5
  c^7}\left(-Q_{kl}^{[3]} x^k \partial_l \mathcal{U}_*+\int \frac{d^3
  \mathbf{y}}{|\mathbf{x}-\mathbf{y}|} Q_{kl}^{[3]} x^k \partial_l
\sigma\right) \\ &= \frac{4 }{5c^{7}} \left(-Q_{kl}^{[3]} x^k \partial_l
\mathcal{U}_*+\mathcal{R}\right)\,,
\end{aligned}
\end{equation}
where the Poisson equation for $\mathcal{R}$ is given in
Eq.~(\ref{eq:PNcorr3}).

Similarly, the expression for the metric correction $(g_{00})_{(9),
  \mathrm{RR}}$ can be written as
\begin{widetext}
\begin{equation}
  \label{eq:g9}
\begin{aligned}
  \left(g_{00}\right)_{(9), {_{\rm RR}}}= & \frac{4}{5 c^9}\left\{-I_{2 k
    l}^{[3]} x^k \partial_l \mathcal{U}_*+Q_{k l}^{[3]}
  x^k\left(-\partial_l \mathcal{U}_2+2 \mathcal{U}_* \partial_l
  \mathcal{U}_*\right)+Q_{k l}^{[4]} x^k\left(-\frac{1}{2} x^l \partial_l
  \mathcal{U}_*+A_{* l}\right)+\frac{5}{126} Q_{k l m}^{[5]} x^k x^l
  \partial_m \mathcal{U}_*\right. \\ & +Q_{k l}^{[5]}
  x^k\left(\frac{17}{42} x^l x^m \partial_m \mathcal{U}_*-\frac{11}{42}
  r^2 \partial_l \mathcal{U}_*\right)-\frac{8}{9} \epsilon_{k l m} S_{m
    n}^{[4]} x^l x^n \partial_k \mathcal{U}_*-2 \mathcal{U}_* \int
  \frac{d^3 \mathbf{y}}{|\mathbf{x}-\mathbf{y}|} Q_{k l}^{[3]} y^k
  \partial_l \sigma \\ & +\int \frac{d^3
    \mathbf{y}}{|\mathbf{x}-\mathbf{y}|}\left[I_{2 k l}^{[3]} y^k
    \partial_l \sigma+Q_{k l}^{[3]} y^k\left[\sigma \partial_l
      \mathcal{U}_*+\partial_l\left(\sigma \delta\right)\right]-3 \sigma
    w_k w_l Q_{k l}^{[3]}-\frac{5}{126} Q_{k l m}^{[5]} y^k y^l
    \partial_m \sigma\right. \\ & +Q_{k l}^{[4]} y^k\left(\frac{1}{2} y^l
    \partial_t \sigma-4 \sigma w_l\right)+Q_{k l}^{[5]}
    y^k\left(-\frac{17}{42} y^l y^m \partial_m
    \sigma+\frac{11}{42}|\mathbf{y}|^2 \partial_l \sigma-\sigma
    y^l\right)-\frac{8}{9} \epsilon_{k l m} S_{m n}^{[4]} y^l y^n
    \partial_k \sigma \\ & \left.\left.-\sigma \int \frac{d^3
      \mathbf{y}^{\prime}}{\left|\mathbf{y}-\mathbf{y}^{\prime}\right|}\left(Q_{k
      l}^{[3]} y^k \partial_l
    \sigma\right)\left[\mathbf{y}^{\prime}\right]\right]\right\}
  \\ \approx & \frac{4 }{5c^{9}} \left(2 Q_{k l}^{[3]} x^k \mathcal{U}_*
  \partial_l \mathcal{U}_*-2 \mathcal{U}_*
  \mathcal{R}+\mathcal{R}_2\right)\,,
\end{aligned}
\end{equation}
\end{widetext}
where, after grouping the terms inside the second integral of~$\int d^3
\mathbf{y}$, a Poisson equation for the potential $\mathcal{R}_2$ is
expressed as [see Eq.~(4.31cc) in Ref.~\cite{Faye2003}]
\begin{widetext}
\begin{equation}
  \label{eq:R2}
\begin{aligned}
\Delta \mathcal{R}_2= & -4 \pi \left[I_2{ }_{k l}^{[3]} x^k \partial_l \sigma+Q_{k
    l}^{[3]} x^k\left[\sigma \partial_l \mathcal{U}_*+\partial_l\left(\sigma
    \delta\right)\right]\right. -3 \sigma w_k w_l Q_{k
    l}^{[3]}-\frac{5}{126} Q_{k l m}^{[5]} x^k x^l \partial_m \sigma \\ &
  +Q_{k l}^{[4]} x^k\left(\frac{1}{2} x^l \partial_t \sigma-4 \sigma
  w_l\right) +Q_{k l}^{[5]} x^k\left(-\frac{17}{42} x^l x^m \partial_m
  \sigma+\frac{11}{42} r^2 \partial_l \sigma-\sigma x^l\right) \\ &
  \left. -\frac{8}{9} \epsilon_{k l m} S_{m n}^{[4]} x^l x^n \partial_k
  \sigma-\sigma \mathcal{R}\right] \\ \approx & -4 \pi \sigma\left[ Q_{k l}^{[3]}
  x^k \partial_l \mathcal{U}_* -3 w_k w_l Q_{k l}^{[3]} - \mathcal{R}\right]\,.
\end{aligned}
\end{equation}
\end{widetext}
We show that the corresponding expression of Eq.~(4.31cc) in Ref.~\cite{Faye2003}
contains a typo in the sign of the term $3 \sigma w_k w_l
Q_{k l}^{[3]}$, as it is easy to deduce when starting from Eq.~(4.18a) in
Ref.~\cite{Faye2003}.

When focusing solely on the leading-order contributions involving
$Q^{[3]}_{ij}$, namely, $\left(g_{00}\right)_{(7), \mathrm{RR}}$ and
$\left(g_{00}\right)_{(9), \mathrm{RR}}$, the corresponding corrections
to the RR metric component, $g_{00, \mathrm{RR}}$, can be expressed as:
\begin{widetext}
  \begin{equation}
    \label{eq:g00}
    \begin{aligned}
      g_{00, {_{\rm RR}}} &\approx \left(g_{00}\right)_{(7), \mathrm{RR}}
      + \left(g_{00}\right)_{(9), \mathrm{RR}} \\ &= \frac{4 }{5} \left[
        \left(-Q_{kl}^{[3]} x^k \partial_l
        \mathcal{U}_*+\mathcal{R}\right) + \left(Q_{k l}^{[3]} x^k
        \mathcal{U}_* \partial_l \mathcal{U}_* -2 \mathcal{U}_*
        \mathcal{R}+\mathcal{R}_2 \right) \right] \\ &= - \frac{4}{5}
      \left[ (1-2 \mathcal{U}_*)\left(Q_{kl}^{[3]} x^k \partial_l
        \mathcal{U}_* - \mathcal{R}\right) \right] +\frac{4}{5}
      \mathcal{R}_2\,,
\end{aligned}
\end{equation}
\end{widetext}
where we note that the second term, \ie $(4/5) \mathcal{R}_2$, is
different from the term reported as $-(8/5) \mathcal{U}_7$ in Eq.~(A.37)
of Ref.~\cite{Oechslin07a}, although our definition of $\mathcal{R}_2$ is
the same as that of $\mathcal{U}_7$ in Ref.~\cite{Oechslin07a} with an
extra term $- 3 Q_{ij}^{[3]} w_i w_j$.

Finally, when computing the GW emission, it is possible to express the
radiating parts of the metric by computing the Newtonian mass quadrupole
formula for the matter distribution and express the second time
derivative of the mass quadrupole $Q_{ij}$ analytically after exploiting
the conservation of rest-mass as~\cite{Blanchet90}
\begin{equation}
  Q_{i j}^{[2]} = \Biggl[2 \int d^3 x D^* \left(V^i V^j+x^i \partial_j
    \mathcal{U}\right)\Biggr]^\mathrm{STF}+\mathcal{O}\left(\frac{1}{c^2}\right)\,.
\end{equation}
The corresponding GW strain at the pole in the two polarisations $+$ and
$\times$ and at a distance $r$ is given by Ref.~\cite{Dimmelmeier05a,Yip2023b}
\begin{align}
  \label{eq:quad_h}
  h_{+} =& \frac{(Q^{[2]}_{xx} - Q^{[2]}_{yy})}{r}\,, \\
  h_{\times} =& \frac{2 Q^{[2]}_{xy}}{r}\,.
\end{align}
%
  
\section{On the frequency of the spacetime update}
\label{sec:cfc_PM_update_freq}

As mentioned in Sec.~\ref{sec:grmhd}, one of the most significant
advantages of the \HO to \BHAC is given by the considerably higher
efficiency with which the evolution can be carried out. Among the various
sources of this efficiency -- which include: a better use of the memory
and AMR, a reduced set of equations for the solution of the field
variables, and a CFL constraint set by the MHD speeds rather than the
speed of light -- the most significant one is probably to be found in the
frequency of the spacetime update. Indeed, in contrast with what happens
in \FIL (but also in all other full-GR MHD codes), the solution of the
field equations needs not be made at every time-level at which the GRMHD
equations are solved. We measure this speed-up in terms of the ratio
$\chi := \Delta t_{\rm met}/\Delta t_{\rm MHD}$, where $\Delta t_{\rm
  met}$ ($\Delta t_{\rm MHD}$) is the time interval between two
successive metric (MHD) updates, and $\chi = 1$ corresponds to the case
in which the field equations are solved at every time-step. We normally
take $\Delta t_{\rm met} \geq \Delta t_{\rm MHD}$ so that $\chi \geq
1$.

\begin{figure*}[ht!]
  \includegraphics[width=0.95\textwidth]{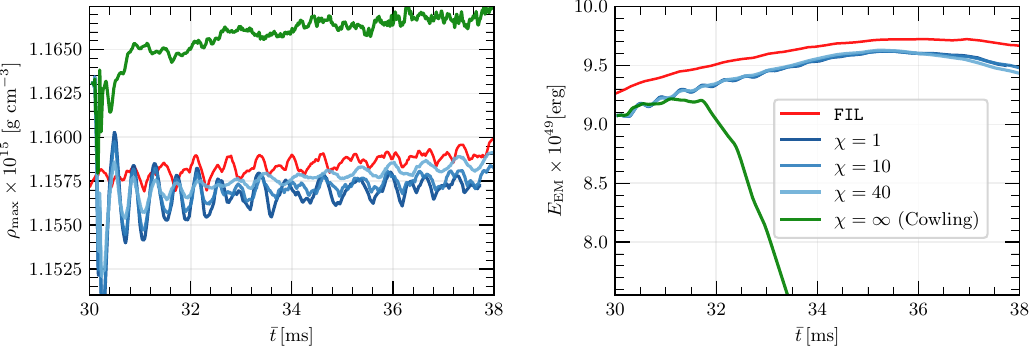}
  \caption{Evolution of the maximum rest-mass density (left panel) and of
    the total EM energy (right panel) as computed by \FIL (red lines) and
    by \BHAC with different metric update frequencies $\chi \geq 1$ (blue
    lines of different shade). The case $\chi=1$ ($\chi=40$) refers to
    when the metric update is done at every ($40$) evolution steps of the
    MHD equations. The data refers to the \HO case \texttt{PM-D} and a
    CFL coefficient $\mathcal{C}_{\rm CFL}=0.2$ has been fixed. The
    green lines show the results relative to a fixed spacetime, \ie
    $\chi=\infty$.}
  \label{fig:different_PM_cfcdt}
\end{figure*}

Figure~\ref{fig:different_PM_cfcdt} shows a comparison in the evolution
of the maximum rest-mass density (left panel) and of the total EM energy
(right panel) between \FIL and \BHAC for the \HO case \texttt{PM-D} (see
also Fig.~\ref{fig:central_dense_post}). It is quite remarkable that the
\BHAC solutions both in terms of rest-mass density and of EM energy are
only weakly dependent on the value of $\chi$ despite that a larger value
leads to a proportionally smaller computational cost. More precisely,
Tab.~\ref{tab:cost} shows the computational costs of two simulations
carried out over one millisecond and having $\chi=1$ ($\Delta t_{\rm met}
= 5.0 \times 10^{-2}~M_{\odot}$) and $\chi=40$ ($\Delta t_{\rm met} =
2.0~M_{\odot}$) differ by a factor $1595/944 \simeq 1.7$; this speedup
becomes of a factor $3117/944 \simeq 3.3$ for $\chi=40$ when comparing
with \FIL. These speedups cannot be attributed to the different AMR
approaches in the two codes, \ie box-in-box in \FIL vs block-based
quadtree-octree AMR in \BHAC, as the two codes have very similar running
times when simulating spherical stars in the Cowling approximation (hence
without the solution of the field equations) and with AMR grid structures
that have the same finest resolution and comparable number of cells.

\begin{table}
  \centering
  \footnotesize
  \setlength{\tabcolsep}{0.8em} 
  \renewcommand{\arraystretch}{1.1}
  \begin{tabular}{lrr}
    \hline
    \hline
    Tests & $\Delta t_{\rm met}$ & ${\rm Cost~for~1~ms}$ \\
             & $[M_{\odot}]$ & $[\rm{CPU~hr}~\rm{ms^{-1}}]$\\
    \hline
      $\chi = 1~~$ (\FIL)  & $1.3\times 10^{-2}$ & $3117$ \\
      $\chi = 1~~$ (\BHAC) & $5.0\times 10^{-2}$ & $1595$ \\
      $\chi = 2~~$ (\BHAC) & $1.0\times 10^{-1}$ & $1222$ \\
      $\chi = 10$~ (\BHAC) & $5.0\times 10^{-1}$ & $1025$ \\
      $\chi = 40$~ (\BHAC) & $2.0$              & $~944$ \\
      $\chi = \infty$ (\BHAC/\FIL) & $\infty$   & $937$/$982$ \\
    \hline
    \hline
  \end{tabular}
  \caption{CPU hours required per millisecond of simulation time in \BHAC
    and the corresponding size of the metric-update time-step for
    different values of $\chi$, as measured with on the \texttt{PM-D}
    \HO. Also reported are the corresponding values for \FIL, for which
    $\chi=1$. Note that all simulations have the same resolution with the
    finest grid spacing of $300\,\rm{m}$.}
  \label{tab:cost}
\end{table}

We should remark that the use of increasingly larger values of $\chi$ has
also the effect of making the evolution slightly more dissipative, as can
be appreciated by considering the amplitude of the post-\HO oscillations,
which are gradually decreased by the use of larger values of $\chi$. At
the same time, the differences between the $\chi=1$ and $\chi=40$ runs
are very small. Indeed, when comparing with the evolution of \FIL, the
$\chi=1$ and $\chi=40$ runs have averaged relative differences of
$-0.3\%$ ($-3.7\%$) and $-0.2\%$ ($-4.0\%$) in the evolution of the
maximum rest-mass density (total EM energy), respectively.

Finally, we note that the computational costs do not reduce significantly
when increasing above $\chi \simeq 40$. Indeed, when omitting altogether
the spacetime update, which corresponds to $\chi = \infty$ (Cowling
approximation), the computational costs are decreased by an additional
$\simeq 0.7\%$. However, as shown in Fig.~\ref{fig:different_PM_cfcdt},
the evolution with a fixed spacetime is extremely inaccurate and is
characterized by much larger central rest-mass densities and a more rapid
decay of the EM energy due to the more efficient braking of the
differential rotation, which causes the suppression of magnetic winding.

\begin{figure*}
\center
\includegraphics[width=0.49\textwidth]{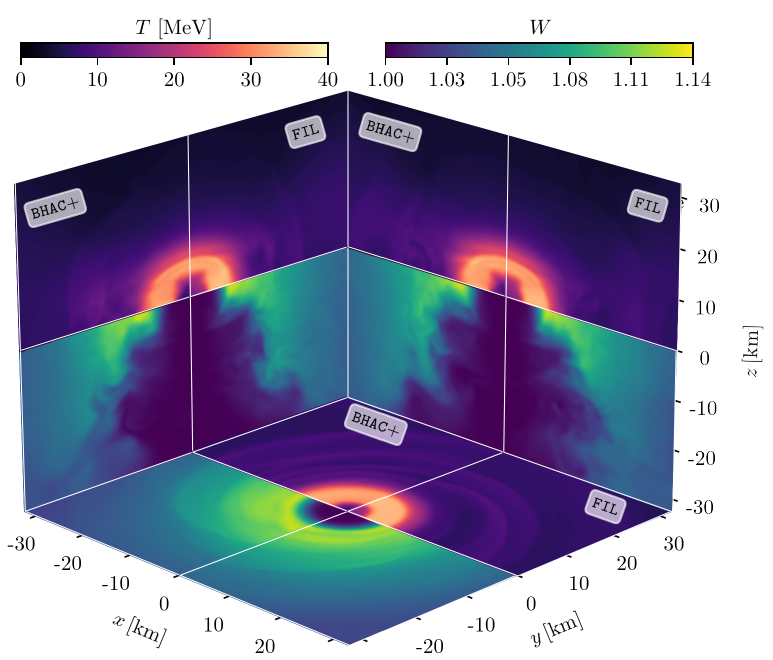}
\hskip  0.25cm
\includegraphics[width=0.49\textwidth]{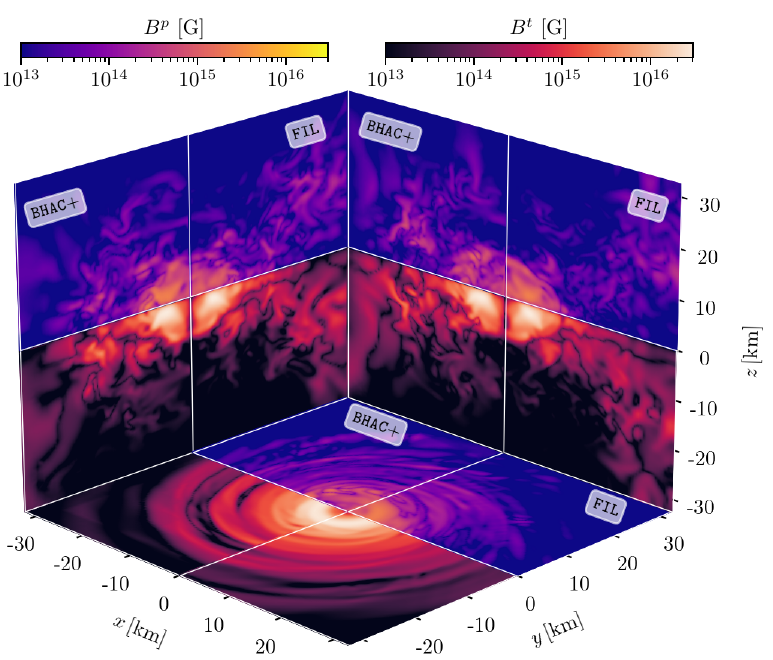}
\caption{The same as in Fig.~\ref{fig:post_comparison_3d}, but for the
  \texttt{PM-C} case, \ie 2D comparison on the principal planes between
  the results of \FIL and \BHAC as computed after $20\, \rm ms$ from the
  \HO of case \texttt{PM-C}.}
\label{fig:post_comparison_3d_PM-C}
\end{figure*}

In view of these tests and considerations, our default implementation is
to use a rather small value of $\chi\simeq 10$ ($\Delta t_{\rm met} =
0.5\, M_{\odot}$) within $\sim 20~\rm{ms}$ after \HO and to increase it
to $\chi \simeq 20-40$ to maximise accuracy and efficiency in the
late-time evolution, when the merger remnant has reached a
quasi-stationary configuration.

\section{Impact on the evolution with an early \HO}
\label{sec:early_ho}

In Fig.~\ref{fig:post_comparison_3d_PM-C}, we compare the results
obtained using two codes, \FIL and \BHAC, after $20\, \rm ms$ of the \HO
process for an earlier \HO case \texttt{PM-C} comparing with the main
text. Although the GW amplitude remains higher in this case compared to
\texttt{PM-D} (as shown in Fig.~\ref{fig:post_comparison_3d}), the
differences between the two codes are quite similar. More specifically,
we observe that key structures, such as the shell structure in
temperature, the ring structure of the Lorentz factor on the equatorial
plane, the wing structure of the Lorentz factor on the $(x,z)$-plane, and
the strong toroidal magnetic field exhibit a good match between the two
codes. Moreover, these prominent structures do not change significantly
relative to the \texttt{PM-D} case, suggesting that the merger remnant is
already well stratified. The poloidal magnetic field appears slightly
more turbulent during this period in both codes, leading to a less
precise match both in terms of magnetic-field topology and of the EM
energy (see Fig.~\ref{fig:central_dense_post}). Overall,
Fig.~\ref{fig:post_comparison_3d_PM-C} reveals that an early \HO is
possible and does not lead to dramatic changes in the dynamics of the
postmerger remnant, especially over timescales $\mathcal{O}(1)\,{\rm
  s}$. At the same time, it provides guidance in preferring, whenever
possible, a \HO that is performed $\sim 30\, \rm ms$ after merger.

\section{On the conformal flatness of BNS postmerger spacetimes}
\label{sec:cy}

The extensive and detailed comparisons between the evolutions from the
full-GR code \FIL and the xCFC code \BHAC discussed in this paper have
provided ample evidence that the CFC approximation provides an accurate
description of the spacetime evolution of the postmerger object. However,
it remains an interesting question to assess and quantify the degree of
``conformal flatness'' of the spacetime generated by the remnant. To this
scope, we recall that the Cotton-York (CY) tensor $\boldsymbol{C}$ is a
rank-2 tensor which consist of third-order derivatives of the spatial
metric $\mathbf{\gamma}$ and is constructed from the more general Cotton
tensor $\boldsymbol{\mathcal{C}}$, which, instead, is of rank three. Its
definition is given by~\cite{Gourgoulhon2007}
\begin{align}
{C}^{ij}:=&-\frac{1}{2}\frac{1}{\sqrt{\gamma}}
\eta^{ikl}\mathcal{C}_{mkl}\gamma^{mj}\\
=&\frac{1}{\sqrt{\gamma}}\eta^{ikl}D_{k}\left(R^{j}_{\phantom{j}l}-
\frac{1}{4}R\delta^{j}_{\phantom{j}l}\right)\,,
\end{align}
where the operator $D_{k}$ denotes the covariant derivative with respect
to the spatial metric. The tensors $R^{j}_{\phantom{j}l}$ and $R$ are the
Ricci tensor and scalar, respectively, with respect to the spatial
metric.

One of the most important properties of the CY tensor is directly
inherited from the Cotton tensor, namely, as the vanishing of the Cotton
tensor $\boldsymbol{\mathcal{C}}$ is a necessary and sufficient condition
for the metric to be conformally flat~\cite{Gourgoulhon2007}, conformally
flat spacetimes admit also a vanishing CY tensor. Hence, following
similar approaches as in Refs.~\cite{Miller04, Ott07b, Iosif2014} we will
make use of this property in order to quantitatively assess how far the
spacetime of merging BNS deviates from being conformally flat.

More specifically, we define the matrix norm of the CY tensor,
$|H_{ij}|$, as the square root of the largest eigenvalue of
$C_{ik}C^{k}_{\phantom{j}j}$ (see Ref.~\cite{Iosif2014} for more
details). In addition, $|H_{ij}|$ is normalised by the covariant
derivatives of $R_{ij}$ in order to provide a local measure of the
deviations from conformal flatness which leads us to the definition
\begin{align}
  \label{eq:H_norm}
  H:=\frac{|H_{ij}|}{\sqrt{D_{i}R_{jk}D^{i}R^{jk}}}\,.
\end{align}
We note that in Ref.~\cite{Iosif2014} it was observed that this
normalization is not suitable for low-density matter distributions
because the denominator $\sqrt{D_{i}R_{jk}D^{i}R^{jk}}$ can exhibit large
variations; fortunately, this drawback has not emerged in our
analysis. Finally, we construct a density-weighted integral using $H$ in
order to obtain an integrated measure over the whole spatial slice
$\Sigma_t $
\begin{align}\label{eq:cyh_int}
\left\langle H \right\rangle _{\rho} :=\frac{\int_{\mathcal{V}}
  \mathrm{d}^3x \sqrt{\gamma} \rho W H}{\int_{\mathcal{V}} \mathrm{d}^3x
  \sqrt{\gamma} \rho W}\,,
\end{align}
such that large/small values of $H$ (and of its time derivative) should
be interpreted as referring to spatial regions with large/small
deviations from conformal flatness. 

\begin{figure*}
  \includegraphics[width=0.96\textwidth]{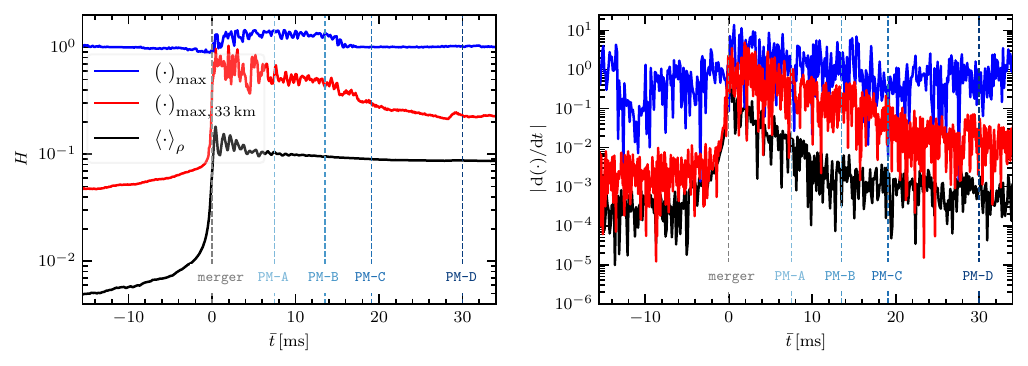}
  \caption{Evolution of various expressions of the normalised measure of
    the deviation from conformal flatness $H$ (left panel; see
    Eq.~\eqref{eq:H_norm}) and of its time derivative (right panel). The
    black lines display the rest-mass density average using
    Eq.~\eqref{eq:cyh_int}, while the blue lines display the maximum of
    $H$ over the entire domain. The red lines visualise the maximum
    value taken over a box centered around $x=y=0$ and having a side
    length of $\sim2\times33~\mathrm{km}$. Shown with vertical blue lines of
    different shades are the different \HO times. 
    }
  \label{fig:cy_evolve}
\end{figure*} 

With these definitions made, we briefly describe in the left panel of
Fig.~\ref{fig:cy_evolve} the evolution of $H$ as a starting point to
evaluate how much the spacetime of the postmerger remnant deviates from
being conformally flat. More specifically, the black line displays the
density average of $H$ using Eq.~\eqref{eq:cyh_int}, while the blue line
reports the maximum of $H$ over the entire domain, and the red line shows
the maximum value taken over the domain inside a box centered around
$x=y=0$ and having a length of $\sim 2\times33~\mathrm{km}$. Note that
with the exception of the maximum of $H$, all other proxies are small,
\eg $\left\langle H \right\rangle_{\rho}\lesssim 10^{-2}$ during the
inspiral, in agreement with the literature on binaries~\cite{Miller04} and
single hot neutron stars formed in the collapse of rotating stellar iron
cores~\cite{Ott07b}. We also note that $(H)_{\mathrm{max}, \, 33\,
  \mathrm{km}}$ shows a significantly larger value than $\left\langle H
\right\rangle_{\rho}$ because the maximum of $H$ is not located inside
either of the two neutron stars.

Furthermore, and as expected, during the merger all three measures
increase rapidly. The quantity $\langle H \rangle_{\rho}$ achieves
moderate values of $\sim 10^{-1}$, which agree with strongly
differentially rotating relativistic stars~\cite{Iosif2014}. More
specifically, these values correspond broadly to the configurations with
the highest values of $T/|W|$ in Ref.~\cite{Iosif2014}, where $T/|W|$
denotes the ratio of the rotational over the gravitational binding
energy; with such values of $T/|W|$, we expect the remnant to still be
stable with respect to the dynamical $\ell =2, m=2$ barmode
instability~\cite{Shibata:2004kb, Baiotti06b, Manca07}, even in the
presence of strong magnetic fields~\cite{Franci2013b}. Between
\texttt{merger} and the time \texttt{PM-A}, all three measures display
strong and rapid variations as a result of the violent
collision-and-bounce cycles of the two stellar cores. After
\texttt{PM-A}, the quantity $\left\langle H \right\rangle_{\rho}$ remains
almost constant as a result of weak GW emission, the dominant mechanism to
change $T/|W|$.

We find that even more informative than the various measurements of $H$
presented in the left panel of Fig.~\ref{fig:cy_evolve} are the
corresponding time-derivatives, which are shown instead on the right
panel. In this case, it is straightforward to realise that those stages
of the process where the spacetime is curved but not highly dynamical,
\ie during the inspiral and at late times in the postmerger, the time
derivatives of $(H)_{\rm max,~33~km}$, and $\langle H \rangle_{\rho}$ are
both very small and either slowly increasing before merger or slowly
decaying after merger. These are indeed the stages when the GW emission
is not very strong. As the binary separation decreases and the merger
takes place, the time derivatives of both proxies follow a behaviour that
is very similar to that associated with the gravitational
radiation. These considerations also highlight a conclusion we had drawn
also before, namely, that a late \HO at about $30\,{\rm ms}$ after merger
is preferred as the GW content in the spacetime is very small. At the
same time, the CFC approach employed with an \HO at about $20\,{\rm ms}$
after merger already presents a very good approximation to the full-GR
spacetime.

\bibliography{aeireferences}

\end{document}